\documentclass[printer]{aa}
\usepackage{graphicx}
\usepackage{txfonts}
\usepackage[colorlinks,citecolor=blue]{hyperref}
\usepackage{threeparttable}

\graphicspath{{./}}

\begin{document}

 \title{Detection and analysis of white-light emission in solar flares through light curve diagnostics}

   \author{Shuanghong Li\inst{1}\fnmsep\inst{2}
        \and Yongliang Song\inst{2}\fnmsep\inst{3}\fnmsep\thanks{Corresponding author: ylsong@nao.cas.cn}
        \and Kun Wang\inst{1}
        \and Xianyong Bai\inst{2}\fnmsep\inst{3}\fnmsep\inst{4}
        \and Xiao Yang\inst{2}\fnmsep\inst{3}
        \and Nian Liu\inst{1}
        \and Ziyao Hu \inst{2}\fnmsep\inst{3}\fnmsep\inst{4}
        }

   \institute{School of Physics and Astronomy, China West Normal University, Nanchong 637009, China
   \and National Astronomical Observatories, Chinese Academy of Sciences, Beijing 100101, China
   \and Key Laboratory of Solar Activity and Space Weather, National Space Science Center, Chinese Academy of Sciences, Beijing 100190, China
   \and School of Astronomy and Space Sciences, University of Chinese Academy of Sciences, Beijing 100049, China}

  \date{Received 04 January 2026 / Accepted 16 March 2026}

  \abstract
   {White-light flares (WLFs) are crucial for understanding the energy transport and heating processes in the lower solar atmosphere. Systematic studies are highly necessary. However, most WLFs are very weak and difficult to detect.}
   {To address this, we propose a new method of detecting WLFs. } 
   {Through the observations of SDO/HMI, the light curve of each pixel in the flaring region can be obtained. By subtracting the slowly varying background, we obtained a series of rapidly varying radiative pulses. Pixels for which radiative pulses during flares significantly exceed those occurring before and after the flare were identified as WL emission regions.}
   {We applied our method to the detection of the X2.2 flare on September 6, 2017 and validated the method. We found that the WL emission in this flare exhibits two phases, and that different regions show distinct WL emission properties. We also detected the WL emission in all the flares (1 X-class, 2 M-class, and 20 C-class) occurred in active region NOAA 12887. It was found that 15 of the 23 flares are WLFs (1 X-class, 2 M-class, and 12 C-class). The occurrence rate of WLFs in this active region is $\sim65\%$. Surprisingly, the occurrence rate of WLFs in C-class flares even reaches up to $60\%$. It should be noted that most of these C-class WLFs are below C5.0. In addition, a C1.0 WLF was identified; this is the lowest GOES-class event with confirmed WL emission to date.}
   {These results demonstrate that WL emission is ubiquitous in most flares, even down to C-class events.}

   \keywords{Sun: activity --
                Sun: flares--
                Sun: photosphere--
                Sun: magnetic fields
               }

   \titlerunning{Detection and analysis of white-light emission in solar flares through light curve diagnostics}
   \authorrunning{Li et al.}

   \maketitle

%===============================================================
\section{Introduction}
\label{sec:intro}

Solar flares are the most violent energy release phenomena on the Sun and are typically observed in $H_\alpha$, ultraviolet (UV), extreme ultraviolet (EUV), X-ray, and radio wavelengths \citep{Fletcher2011}. Since the energy release in solar flares predominantly occurs in the corona, it is difficult for the energy to propagate downward to the lower atmospheric layers, such as the photosphere. Only under extremely special circumstances can the lower atmosphere be effectively heated, thereby generating significant white-light (WL) continuum emissions \citep{Neidig1983, Neidig1989}. Flares exhibiting such obvious enhancement in the visible continuum spectrum are then termed white-light flares (WLFs). The first WLF was observed by \citet{Carrington1859} and \citet{Hodgson1859} on September 1, 1859. Up to the present, only a small fraction of solar flares have been identified as WLFs \citep[e.g.][]{Matthews2003, Fang2013, Song2018b, Jing2024, Cai2024, Li2024}.

The key question regarding WLFs is how the energies transport to and heat the lower solar atmosphere \citep[e.g.][]{Neidig1989, Ding1999b}. Various mechanisms have been proposed, such as non-thermal electron beam bombardment heating \citep[e.g.][]{Hudson1972, Krucker2015, Watanabe2020}, Alfv\'en wave heating \citep[e.g.][]{Emslie1982, Fletcher2008, Xuzhe2025}, chromospheric condensation \citep[e.g.][]{Gan1992, Kowalski2015}, and radiative backwarming \citep[e.g.][]{Machado1989, Metcalf1990, Alatalo2014}. However, it is difficult to distinguish which mechanism plays a dominant role in a WLF due to the limited observations. Some studies have proposed  that multiple heating mechanisms may coexist and play significant roles simultaneously in the same WLFs \citep[e.g.][]{Xu2010, Hao2017, Song2023}.

According to the observations, WLFs can be divided into two types \citep{machado1986, Fang1995, ding1999a,Ding1999b}. Type-I WLFs have a good temporal and spatial correspondence between the maximum of the continuum spectrum radiation and the maximum of the hard X-ray (HXR) and radio radiation. There is a strong Balmer jump in the spectra, and the Balmer lines are very broad and strong \citep{Neidig1984, Fang1995}. For type-II WLFs, on the contrary, the continuum radiation maxima do not have a strong correspondence with the HXR and radio radiation maxima. There is no or only a very weak Balmer jump in the spectra. The Balmer lines are also weak and narrow \citep{ding1999a,Ding1999b}. The difference between the two types suggests that WLFs may have different physical mechanisms and origins \citep{Fang1995}. However, up to the present observation, most of the WLFs belong to type I and only a few belong to type II.

Recently, magnetic field structure in an active region was thought to play a significant role in producing WLFs, especially for these middle- and low-class flares \citep{Song2018a, Song2018b, Song2018c, Song2020}. Among the 70 flares of the active region NOAA 11515 during its passage across the solar disc, \citet{Song2018b} found that 20 of them are WLFs, occurred mainly on two days, and are located along a narrow ribbon with the negative magnetic field surrounded by the positive field on both sides. In a statistical study, \citet{Song2018c} found the occurrence rate of WLFs for circular-ribbon flares greater than M1.0 is nearly 70\%, implying that the fan-spine magnetic field topology tends to produce WLFs. This should be due to the interaction between the strong explosive structure (flux rope) and the strong confined structure (fan-like background field), and the released energy reacts back to the lower solar atmosphere \citep{Song2018a}. In the case study of a small compact C2.0 WLF, \citet{Song2020} found that the reconnection site lies right at the centre of a U-shaped magnetic field structure and concluded that this WLF was powered by the low-height magnetic reconnection.

The successful launch of a series of space solar telescopes, such as the Solar Dynamics Observatory \citep[SDO;][]{Pesnell2011} and the Advanced Space Solar Observatory \citep[ASO-S;][]{Gan2023}, provides us an unprecedented opportunity to study WLFs systematically. As the first step, we need to find out a sufficient number of WLFs. However, most WLFs show very weak WL enhancements and it is a great challenge to detect them. Only a select number of flares -- such as the Carrington event -- can release vast amounts of energy, efficiently heating the lower atmosphere and producing exceptionally bright WL radiation compared to quiet regions. These WLFs can even be identified from a single snapshot during the flare. However, for the majority of WLFs, the WL enhancements are not easily  recognized from a single static image. We need to compare the WL images at different times.  A traditional method is to make a movie and identify the WL enhancements by checking the rapid evolutionary changes. To enhance the visibility of some very weak WL enhancements, \citet{Song2018b} made the movie using the pseudo-intensity images constructed by magnifying the difference between two adjacent WL images. Nevertheless, the movie-based detection of WLFs requires subjective human intervention and is unsuitable for large-sample WLF identifications.

The more prevalent method is using the difference maps between two times, i.e. $\Delta I=I_p-I_0$. Here $I_p$ and $I_0$ refer to the continuum images at the peak time of the flare and the time before the flare, respectively. By setting a threshold, the WL enhancement regions can be delineated. However, the peak times of WL emissions for different pixels could be different. Therefore, when we set the same peak time and pre-flare time for all the pixels, some WL enhancement regions may be missed, because their brightening peak times are earlier or later than the peak time we set, and the $\Delta I$ calculated for these regions may fall below the threshold for WLF.

We therefore need to check the WL emission at each pixel during the flare. \citet{Cai2024} proposed an optimized WLF identification method. They first made running difference images using the HMI continuum intensity maps, and then set the intrinsic threshold for each pixel by calculating the WL emssion fluctuation at the pixel.  When the running difference values ($\delta_n$) at a pixel are larger than the intrinsic threshold by three times or more, this pixel would be identified as a valid WL enhancement region. However, as we all know, magnetic field structure of an active region could be changed after flare and the corresponding sunspot intensity would also be changed permanently \citep[e.g.][]{Dengna2005, Wanghaimin2010, Wangshuo2012, Song2016}. This permanent WL intensity change induced by magnetic field restructuring, rather than flare energy deposition, may be misinterpreted as true WLF emission under their detection methodology.

In this paper, we propose a new method of detecting the WLFs by analysing the continuum light curve at each pixel in the flare region. The paper is organized as follows. The data and observations are given in Sect. \ref{Sec2} and we describe our method in Sect. \ref{Sec3}. A detailed study on the X2.2 flare on September 6, 2017 based on our method is given in Sect. \ref{Sec4}. Then we detect the WLFs in an active region of NOAA 12887 and give the result in Sect. \ref{Sec5}. At last, we give a brief summary and discussion in Sect. \ref{Sec6}.

%===============================================================
\section{Observations and data reduction}
\label{Sec2}

The data used in this study are mainly from the Atmospheric Imaging Assembly \citep[AIA;][]{Lemen2011} and the Helioseismic and Magnetic Imager \citep[HMI;][]{Scherrer2011} on board the SDO. The AIA provides the higher solar atmosphere images with a spatial resolution of about $1^{\prime\prime}$ and a pixel size of about $0.6^{\prime\prime}$. The temporal resolutions for EUV passbands (94, 131, 171, 193, 211, 304, 335 \AA) and UV passbands (1600 and 1700 \AA) are 12 s and 24 s, receptively. In this work, we used the 1600 Å images to identify the flare ribbons. The HMI provides the continuum intensity maps, Dopplergrams, line-of-sight (LOS) magnetograms, and vector magnetograms by using the spectral line of Fe {\scriptsize I} 6173 \AA, with a spatial resolution of about $1^{\prime\prime}$ and a pixel size of about $0.5^{\prime\prime}$. The corresponding temporal resolutions are 45 s, 45 s, 45s, and 720s. In this study, we mainly used the continuum intensity maps, LOS magnetograms, and vector magnetic fields. We also used the 1 - 8 \AA\ soft X-ray (SXR) data observed by the Geostationary Orbiting Environmental Satellites \citep[GOES;][]{Hanser1996} to show the temporal evolution of flare.

\begin{figure}[!ht]
\centerline{\includegraphics[trim=1.0cm 0.5cm 0.5cm 1.0cm, width=0.47\textwidth]{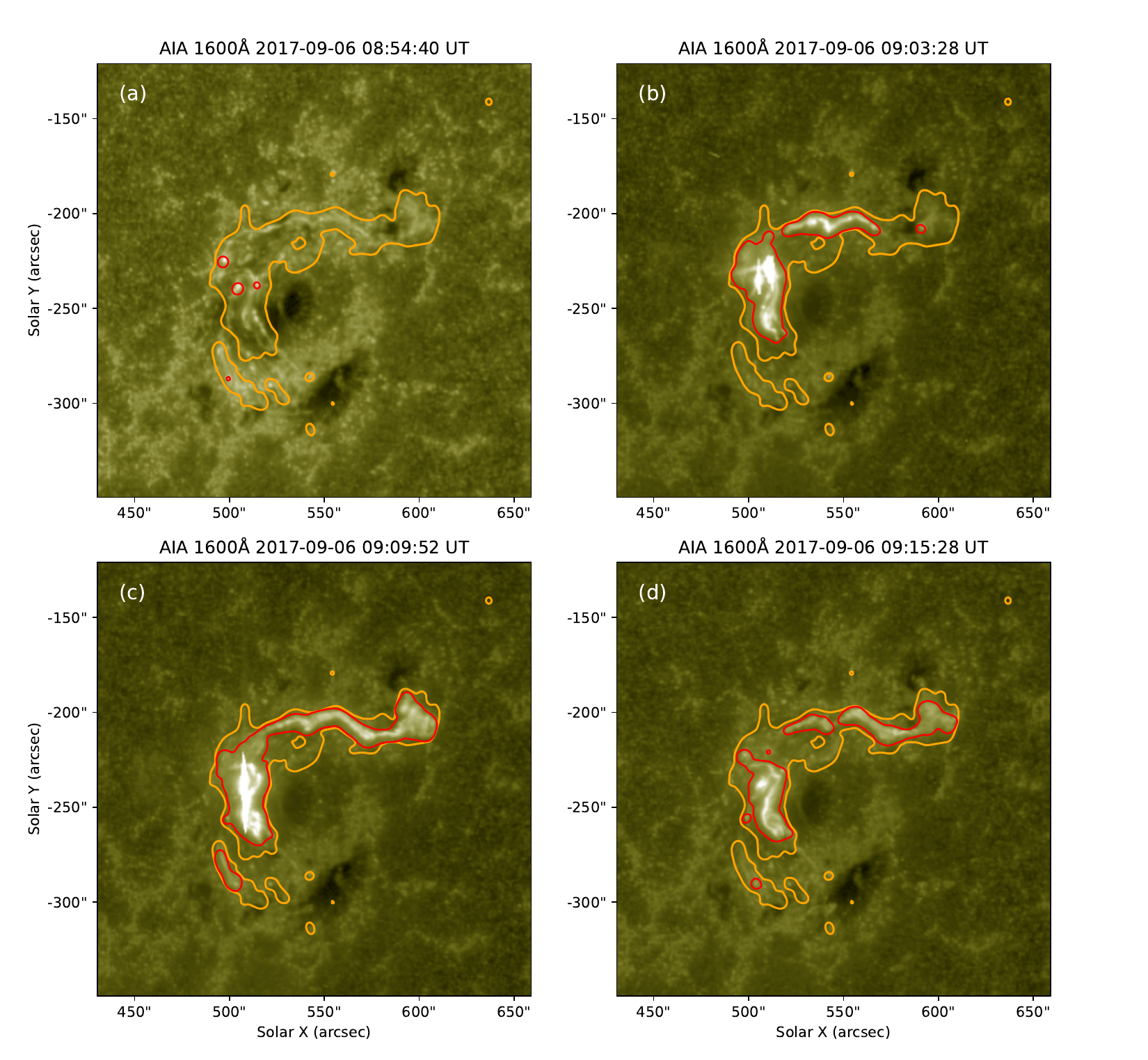}}
\caption{AIA 1600 Å images showing the X2.2 flare ribbons at different times on September 6, 2017. The red contours represent the flare brightening regions identified at each time step, while the orange contours indicate the cumulative brightening region of the flare across all times. The contour level is $(I_{1600f}-I_{1600q})>20\sigma_{1600q}$. Here $I_{1600f}$ and $I_{1600q}$ refer to the AIA 1600 Å intensities in the flare region and quiet region, respectively. $\sigma_{1600q}$ is the standard deviation of AIA 1600 Å intensities in a $100^{\prime\prime}\times100^{\prime\prime}$ quiet region.}
\label{fig1}
\end{figure}

The data processed mainly using the SunPy\footnote{\url{https://sunpy.org}}. We performed data alignment and standardized the time format for different data. We first show our method by analysing the X2.2 flare on September 6, 2017 and then give some interesting results. We further validate our method's effectiveness by applying it to detect WLFs in an active region of NOAA 12887, which produce 23 flares during its passage across the solar disc, including 1 X- and 2 M- and 20 C-class flares.

%===============================================================
\section{WLF identification via light curve diagnostics} 
\label{Sec3} 

The X2.2 WLF on September 6, 2017 occurred in the very well-known active region of NOAA 12673. This active region underwent a rapid and complex magnetic field evolution and ultimately produced the largest flare (X9.3) of Solar Cycle 24 \citep[e.g.][]{Yang2017}. We selected this X2.2 flare for methodological study due to its position relatively far away from the solar limb. And this flare is a confined flare, with the energy predominantly released locally and without triggering a large-scale coronal mass ejection (CME). In addition, this flare clearly exhibits a two-phase evolution and displays rich dynamical details \citep[e.g.][]{Liu2018,Gopasyuk2022}. We first identified the flaring region, then performed light-curve diagnostics for all the pixels within the region, and finally determined the location of WL enhancement, the magnitude of enhancement, and other associated physical information.
%------------------------------------------------------------------
\subsection{Identifying the flaring region} \label{Sec3.1}

Figure \ref{fig1} shows the AIA 1600 \AA\ images at different times for the X2.2 flare. The red and orange contours represent the flaring regions identified at each time step and the cumulative flaring region across all times, respectively. We defined the flaring region with $(I_{1600f}-I_{1600q})>20\sigma_{1600q}$. Here $I_{1600f}$ and $I_{1600q}$ refer to the AIA 1600 Å intensities in the flaring region and quiet region, respectively. $\sigma_{1600q}$ is the standard deviation of AIA 1600 Å intensities in a $100^{\prime\prime}\times100^{\prime\prime}$ quiet region. It is well known that WLFs result from the flare energy deposition into the lower atmosphere, leading to localized heating. Therefore, we infer that the WL enhancement regions should be located within the flaring areas, and then we mainly focus on the characteristics of the light curves of all the pixels within the flaring region.

%------------------------------------------------------------------
\subsection{Analysis of light curves} \label{Sec3.2}

\begin{figure}[!ht]
\centerline{\includegraphics[trim=0.0cm 0.5cm 0.0cm 0.0cm, width=0.52\textwidth]{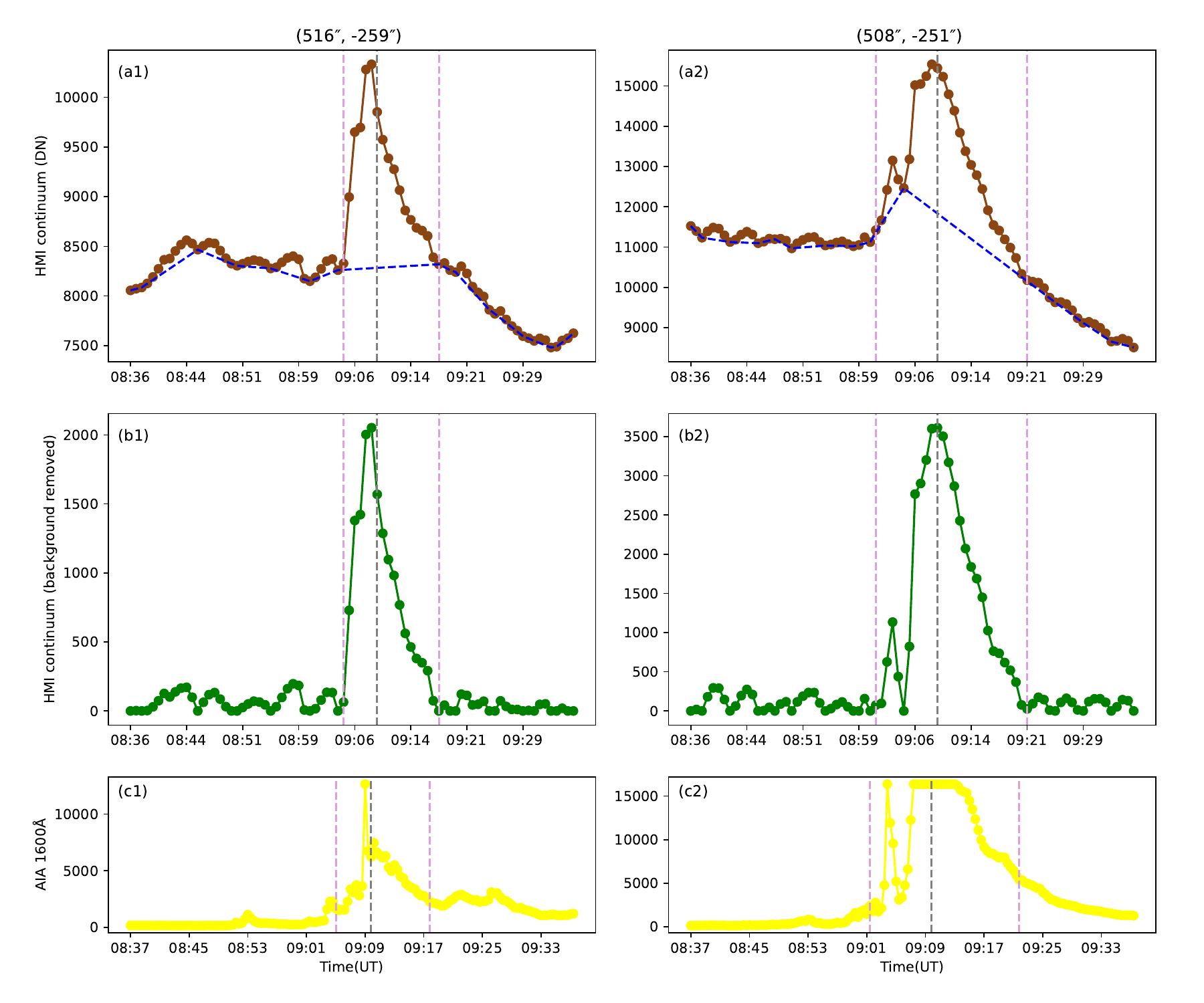}}
\caption{Temporal evolutions of HMI continuum and AIA 1600 \AA\ intensities at two positions in WL enhancement regions. (a1) and (a2): HMI continuum light curves for the two WL positions, respectively. The dashed blue line indicates the background trend of the pixel. (b1) and (b2): HMI continuum light curves after removing the background, which exhibit a series of WL emission pulses. (c1) and (c2): AIA 1600 Å light curves at these two positions, respectively. The vertical grey line marks the peak time of the flare, and the vertical purple lines mark the beginning and end times of the 1600 Å impulsive emission in this pixel.}
\label{fig2}
\end{figure}

\begin{figure}[!ht]
\centerline{\includegraphics[trim=0.0cm 0.5cm 0.0cm 0.0cm, width=0.52\textwidth]{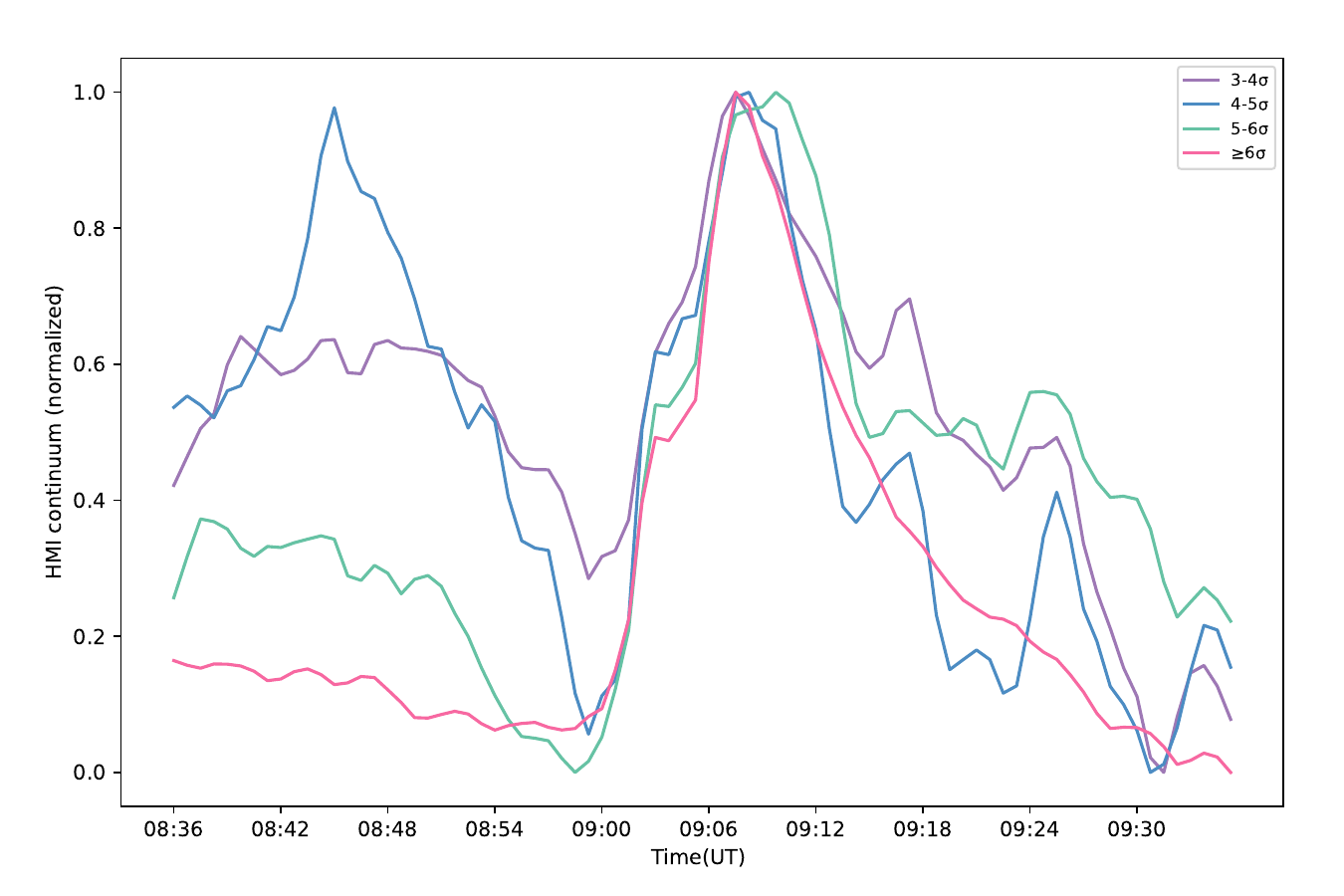}}
\caption{Averages of HMI continuum light curves (normalized) within different threshold ranges. The threshold ranges are $3\sigma<(I_{fm}-\frac{1}{N}\sum_{i=1}^{N} I_{0i})<4\sigma$ (purple), 
$4\sigma<(I_{fm}-\frac{1}{N}\sum_{i=1}^{N} I_{0i})<5\sigma$ (sky blue), $5\sigma<(I_{fm}-\frac{1}{N}\sum_{i=1}^{N} I_{0i})<6\sigma$ (green), $(I_{fm}-\frac{1}{N}\sum_{i=1}^{N} I_{0i})\ge6\sigma$ (pink). Here $I_{fm}$ refers to the maximum value of all the WL emission pulses during the flare (see Figs. \ref{fig2}b1 and \ref{fig2}b2). $I_{0i}$ refer to the values of WL emission pulses before and after the flare (also see Figs. \ref{fig2}b1 and \ref{fig2}b2). And $\sigma$ is the corresponding standard deviation for $I_{0i}$.}
\label{fig3}
\end{figure}

We plot the HMI continuum light curves for each pixel within the flaring region. The start and end times of the curves are set as 10 minutes before the flare occurrence and 10 minutes after the flare ends. Figures \ref{fig2}a1 and \ref{fig2}a2 show the HMI continuum light curves at two positions. Such light curves can be divided into two parts: one is a slow and even quasi-periodic variation, mainly originating from the p-mode waves, granulation motions, or gradual magnetic field evolution; the other is a rapid pulse-like variation that is speculated to be related to the instantaneous heating effect from, for example, the flares, jets, or other energy-released activities. Given that WLF is fundamentally a heating phenomenon in the lower solar atmosphere, its corresponding feature should be the latter, i.e. the rapid pulse-like variation.

Based on this idea, we decomposed the HMI continuum light curve at a pixel into two parts: one is the slow and even quasi-periodic background variation; the other is the pulse-like variation superimposed on it. The specific processing method is as follows: we identified a series of minimum points on the light curve and connected these points to form a background variation curve (dashed blue lines in Figs. \ref{fig2}a1 and \ref{fig2}a2); after subtracting this background from the original curve, a series of WL radiation pulses can be obtained (see Figs. \ref{fig2}b1 and \ref{fig2}b2).

The judgment criteria for WLF light curves are as follows: when the maximum radiation pulse appearing during the flare has an intensity greater than a certain threshold decided by the fluctuation amplitude of the radiation pulses before and after the flare, then the light curve is regarded as a WLF emission curve, and the corresponding pixel is determined as a WL point. Section \ref{Sec3.3} details the threshold criteria established for our analysis.
%------------------------------------------------------------------
\subsection{Setting the threshold} \label{Sec3.3}

After subtracting the background from the HMI continuum light curve at a pixel within the flare region, we could obtain a series of WL radiation pulses at that pixel (Figs. \ref{fig2}b1 and \ref{fig2}b2). The threshold was determined based on the fluctuation amplitude of these WL radiation pulses before and after the flare. It should be noted that the pre-flare and post-flare baselines were determined pixel by pixel based on the AIA 1600 \AA\ flux at each location (Figs. \ref{fig2}c1 and \ref{fig2}c2). We calculated the mean value and standard deviation ($\sigma$) of these WL radiation pulses that appeared before and after the flare. Given that a flare is a violent energy release process, its heating efficiency on the atmosphere is much higher than that of random heating processes. Therefore, we infer that the value of $(I_{fm}-\frac{1}{N}\sum_{i=1}^{N} I_{0i})$ should be at least greater than $3\sigma$. Here $I_{fm}$ refers to the maximum value of the WL emission pulses during the flare. $I_{0i}$ refer to the values of these WL emission pulses before and after the flare. $N$ is the number of the WL emission pulses appeared  before and after the flare. The intensities of all the WL emission pulses correspond to a series of peaks in the residual intensity after background subtraction.

Figure \ref{fig3} shows the normalized mean values of light curves for all the pixels within different threshold intervals. We see that the higher the threshold that is set, the more significant the difference between the peak during the flare and the pulse peaks before and after the flare. When the threshold is set to $6\sigma$, the overall curve shows an obvious flat feature before the flare. Based on the above analysis, we finally determined the threshold as $6\sigma$, i.e. $(I_{fm}-\frac{1}{N}\sum_{i=1}^{N} I_{0i})\ge6\sigma$.

%------------------------------------------------------------------
\subsection{Isolated WL pixels} \label{Sec3.4}

\begin{figure}[!ht]
\centerline{\includegraphics[trim=0.0cm 0.7cm 0.0cm 0.0cm, width=0.52\textwidth]{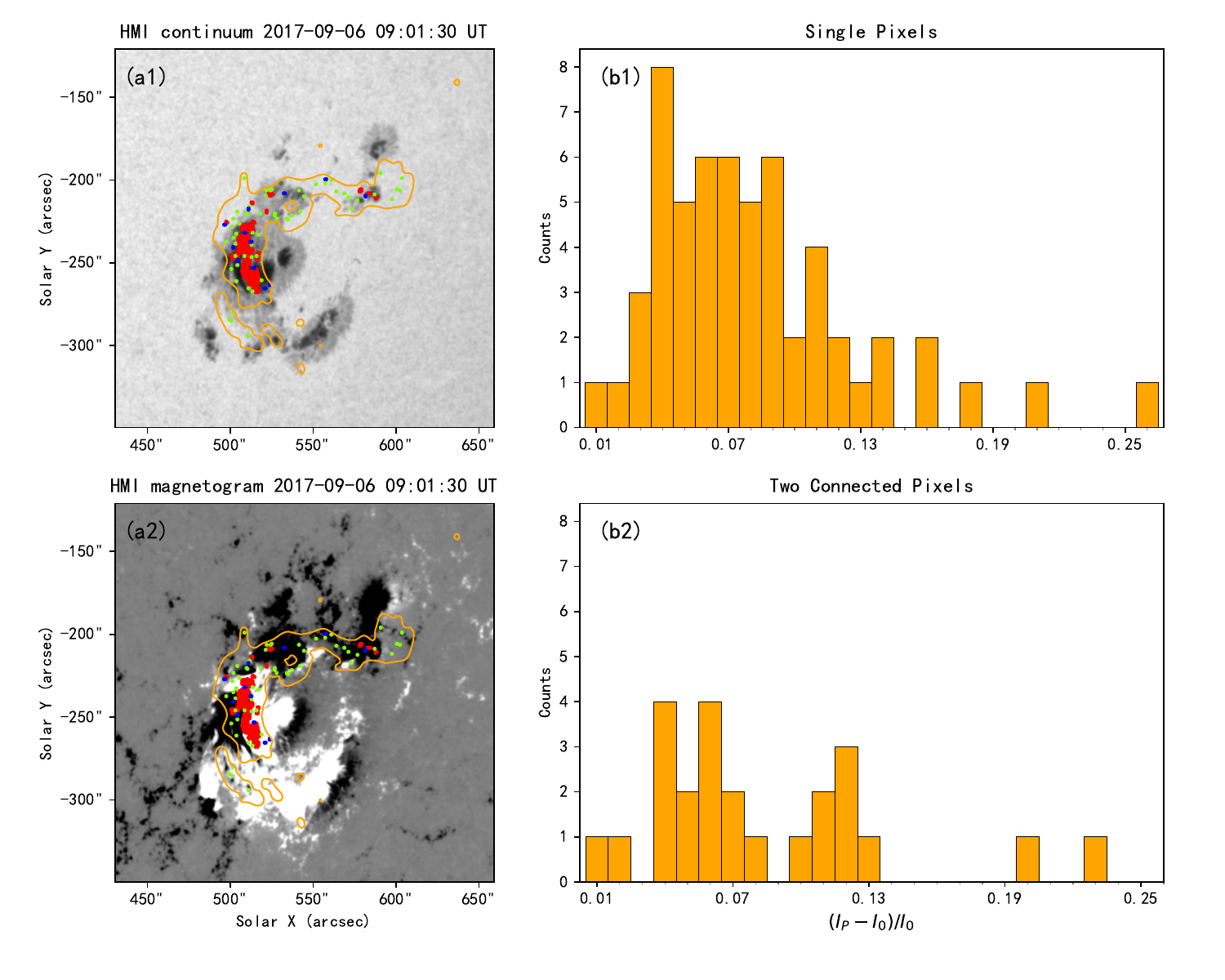}}
\caption{(a1) and (a2): HMI continuum image and LOS magnetogram for the active region NOAA 12673 at 09:01:30 UT on September 6, 2017. The orange contours mark the flare regions. The red region indicates the area of concentrated WL pixels. The green denotes a single isolated WL pixel, while the blue denotes only two conjoined WL pixels.  (b1) and (b2): Statistical plots of the WL enhancement ($(I_p-I_0)/I_0$) for the isolated single-pixel region and the adjacent two-pixel region, respectively. Here $I_p$ and $I_0$ refer to the HMI continuum intensities at the peak of WL flux and before the flare at a pixel.}
\label{fig4}
\end{figure}

\begin{figure}[!ht]
\centerline{\includegraphics[trim=0.0cm 0.5cm 0.0cm 0.0cm, width=0.53\textwidth]{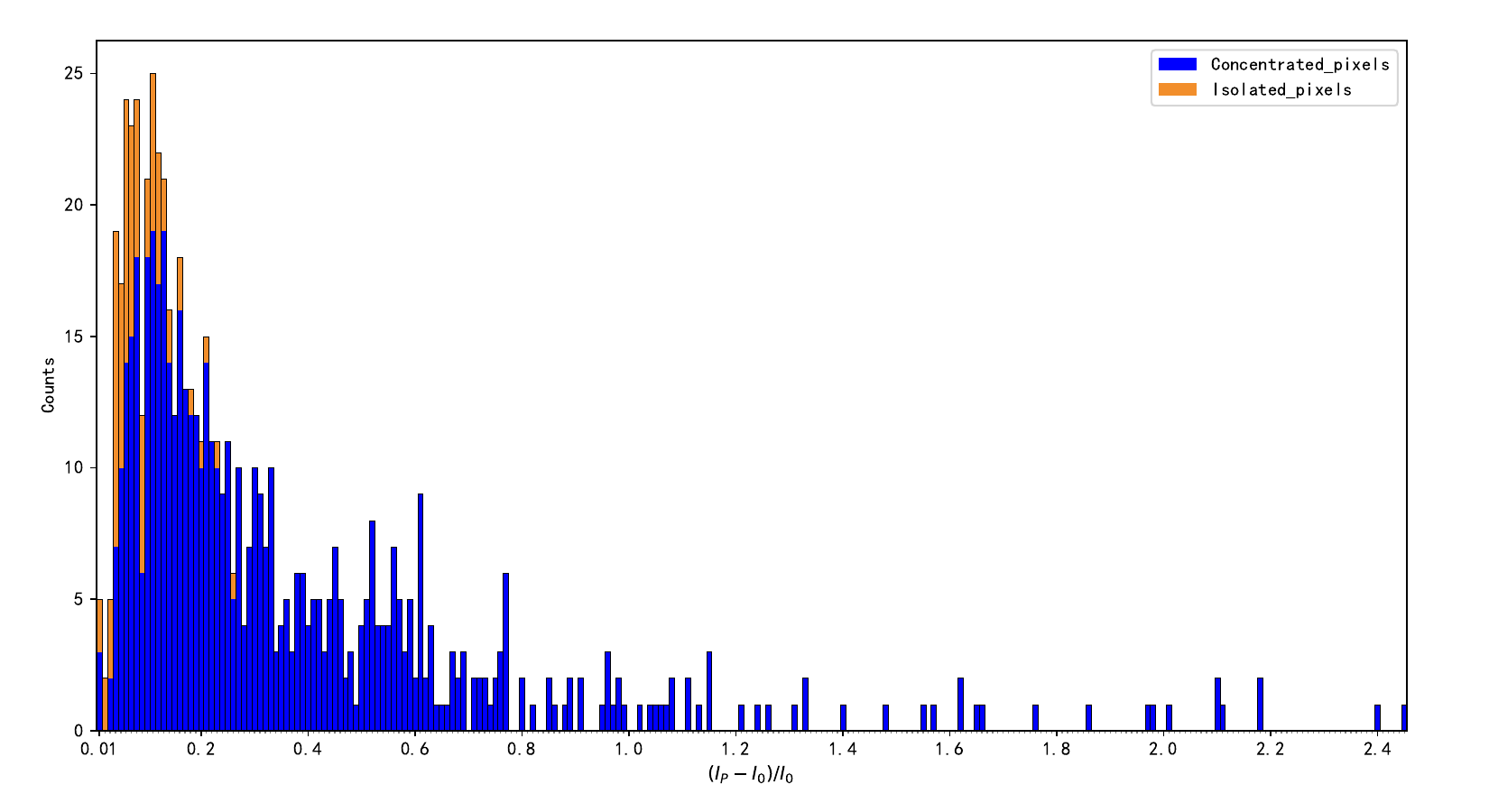}}
\caption{Statistical plot of the WL enhancement ($(I_p-I_0)/I_0$) for contiguous and isolated WL pixels. These pixels are shown in Fig. \ref{fig4}.}
\label{fig5}
\end{figure}

Among the WL pixels identified based on above method, we find that some of them are isolated. Figure \ref{fig4} presents the distributions and statistical results for the single isolated WL pixels and two adjacent isolated WL pixels. Both of them are scattered across the entire flare ribbon, with a significant fraction located in network field and quiet-Sun regions, whereas non-isolated points are concentrated in the flare's core region, i.e. around the polarity inversion line (PIL) areas (see panels a1 and a2).

From panels (b1) and (b2), we see that the number of two adjacent isolated WL pixels is about half that of single isolated WL pixels. Their corresponding WL enhancement magnitudes are basically similar. The medians of WL enhancements, i.e. $(I_p-I_0)/I_0$, for both of them are all approximately $\sim7\%$. Here $I_p$ and $I_0$ refer to the HMI continuum intensities at the peak of WL flux and before the flare. Interestingly, \citet{Hudson2006} studied 11 WLFs in the WL channel of TRACE and gave the minimum excess contrast as only $0.08 \pm 0.017$. \citet{Kuhar2016} studied 43 WLFs (X- and M-class) and gave the lowest WL enhancement as $0.08 \pm 0.07$.  \citet{Song2018b} studied the WLFs occurred in NOAA 11515 and found that the lowest value for WL enhancement is $8\% \pm 1.3\%$. This indicates that these isolated WL pixels exhibit very weak WL emissions. Since no significant difference exist between single isolated WL pixels and two adjacent isolated WL pixels, we unify them as `isolated WL pixels'.

Figure \ref{fig5} compares these isolated WL pixels with non-isolated WL pixels. We see that the number of isolated WL pixels is very small compared with the non-isolated WL pixels: about 10\% of the number of  non-isolated WL pixels. The WL enhancements for most isolated WL pixels are lower than 10\%, much lower than for the most non-isolated WL pixels. Obviously, these isolated WL pixels represent regions with weak enhancements, whereas the non-isolated ones correspond to regions with strong WL emissions. Considering that flares generally occur above PIL region, and to exclude some random and accidental heating processes, we excluded these isolated WL pixels.

In this X2.2 WLF, the largest WL enhancement for isolated WL pixels reaches only $\sim$25\%, placing it at the middle-to-lower level among all the non-isolated WL points. Such a value is significantly weaker than the extremely small-scale WLF reported by \citet{Jess2008}, which exhibited a WL enhancement of $\sim$300\% and a kernel size of only $\sim$300 km. Therefore, we consider it relatively safe to exclude these isolated WL points from our analysis. However, we must be very careful to treat isolated WL points with exceptionally strong enhancements (e.g. >200\%) in future works. In particular, only several isolated points show strong WL radiation without significant WL emission from other concentrated regions, which is very similar to the event reported by \citet{Jess2008}. Potential artefacts such as cosmic-ray hits or CCD hot pixels must be carefully ruled out. Confirmation should also rely on supplementary continuum or spectroscopic observations from other instruments to avoid missing extreme and atypical WLF events. This further underscores the importance of high-resolution observations for accurate WLF diagnosis.
%------------------------------------------------------------------
\subsection{Advantages of this WLF recognition method} \label{Sec3.5}

\begin{figure}[!ht]
\centerline{\includegraphics[trim=0.0cm 0.8cm 0.0cm 0.0cm, width=0.55\textwidth]{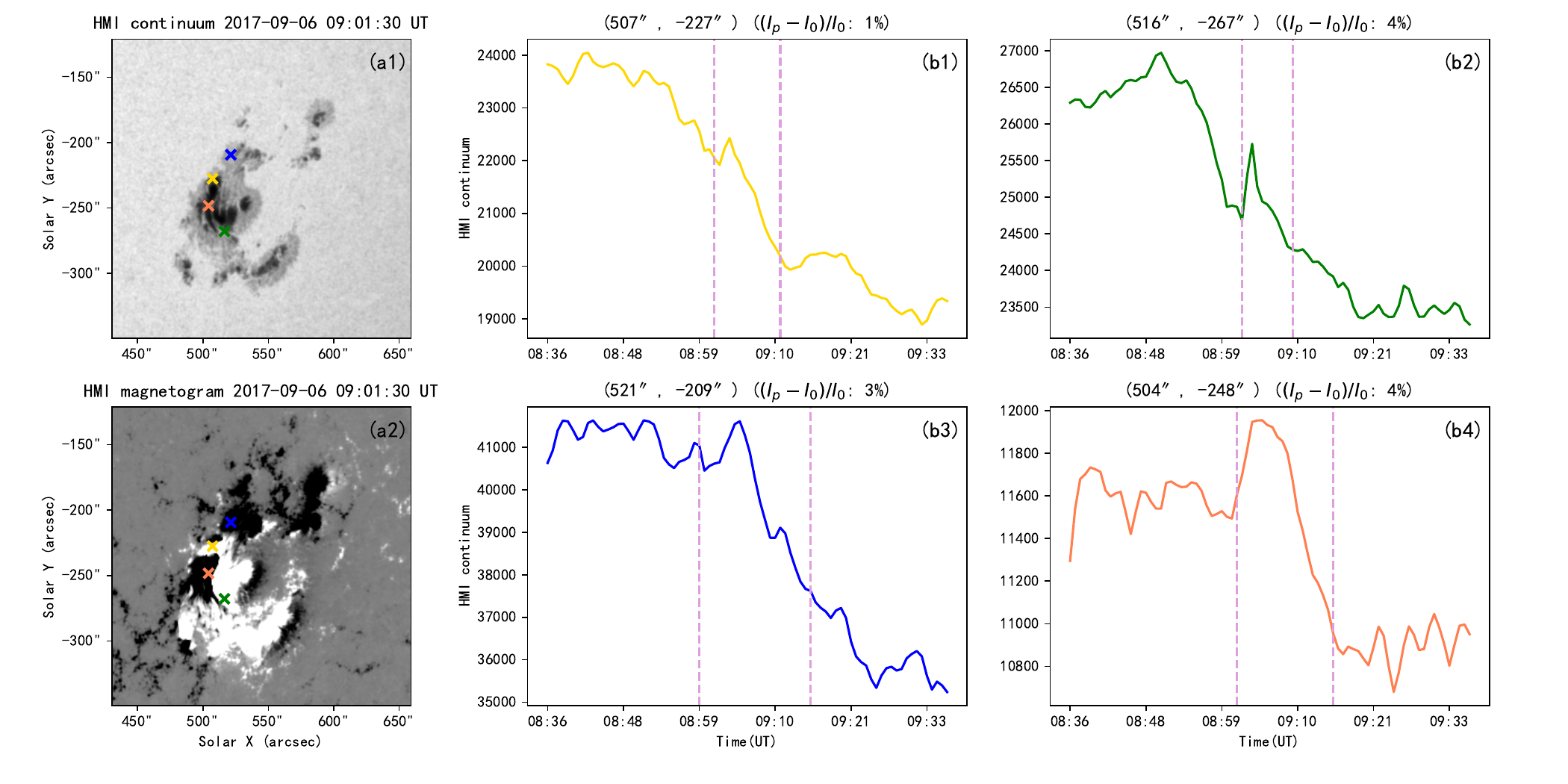}}
\caption{HMI continuum light curves showing very weak WL emissions during the flare at four different positions. (a1) and (a2): HMI continuum map and LOS magnetogram of the NOAA 12673 at the time of 09:09:45 UT on September 6, 2017.  The four `$\times$' symbols in different colours mark the positions where we show the HMI continuum light curves.  (b1)-(b4): Corresponding HMI continuum light curves at four different positions. The vertical purple lines mark the beginning and end times of the 1600 Å impulsive emission at the position. The corresponding WL enhancements ($(I_p-I_0)/I_0$) are given at the top of these panels.}
\label{fig6}
\end{figure}

Traditionally, using the difference imaging technique, people need to set a lower limit for WL enhancement, such as $(I_p - I_0)/I_0 \ge 5\%$ or $(I_p - I_0)/I_0 \ge 8\%$ \citep[e.g.][]{Song2018b, Song2018c, Jing2024, Li2024}. However, our method does not require a specially set lower limit for WL enhancement. It can detect extremely weak enhancement signals. Figure \ref{fig6} shows the HMI continuum light curves at four different pixels. We see the minimum WL enhancement is only 1\%, and the pulse-like WL radiation signal during the flare can be clearly seen from the light curve. The WL enhancements for the other three pixels are 3\%, 4\%, and 4\%.  Undoubtedly, our detection method demonstrates a better detection capability compared with the traditional ones.

\begin{figure}[!ht]
\centerline{\includegraphics[trim=0.0cm 0.8cm 0.0cm 0.0cm, width=0.53\textwidth]{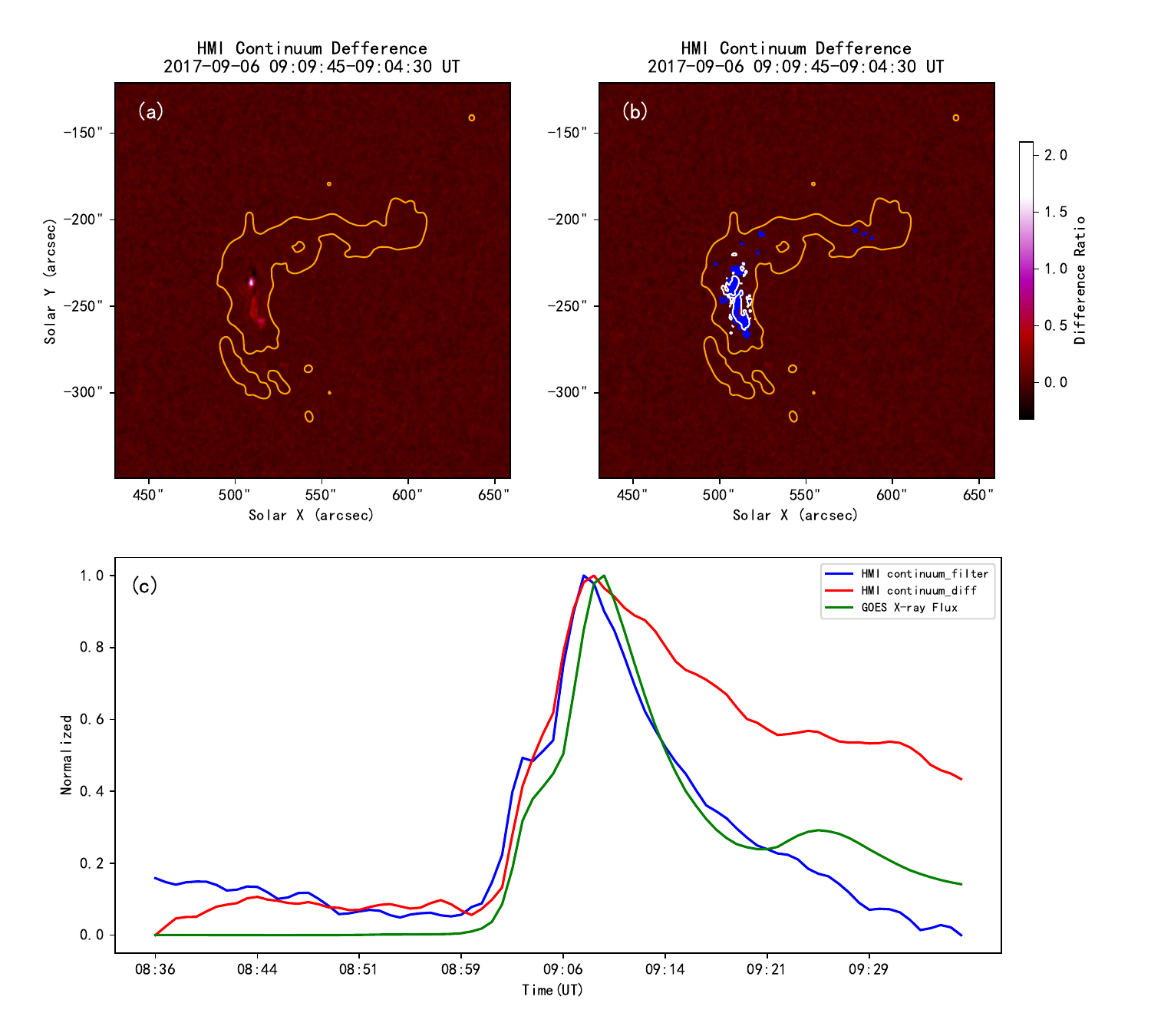}}
\caption{Comparison of two WLF identification methods. (a) and (b): HMI continuum difference image ($(I_p-I_0)/I_0$). The orange contours mark the flare region. The blue regions represent the WL pixels identified by our method, and the white contours indicate the regions where the difference is greater than 8\%.  (c): Normalized average HMI continuum light curves for the WL emission regions identified by different methods. The blue light curve was obtained with our method. The red light curve was obtained based on the difference image. The green light curve indicates the GOES soft-X ray (SXR) 1-8 \AA\ flux. }
\label{fig7}
\end{figure}

The traditional method using the difference image ($I_p - I_0$) calculated between two times in a flare may miss some WL enhancement regions. After all, the WL radiation variations at different positions in the flaring region are not necessarily synchronous. However, our method does not encounter such issues. It independently analyses and verifies the light curve at each pixel in the flaring region. Figures \ref{fig7}a and \ref{fig7}b show the WL pixels identified by different methods. Obviously, our method identifies more WL pixels. 

Figure \ref{fig7}c shows the normalized average of the continuum light curves obtained by the two methods. The WL curve obtained by our method shows obvious pulse-like characteristics, while the WL curve obtained by the traditional difference imaging method exhibits an exceptionally gradual decline in the flare decay phase. The corresponding values at the decay phase are much higher than those before the flare, which is presumably affected by these permanent WL emission changes caused by the permanent magnetic field structure changes after the flare. Obviously, our method is unaffected by permanent WL emission changes in some areas during the flare. 

Moreover, the WL curve obtained by our method clearly exhibits two stages during the flare rising phase, which is consistent with the trend of the GOES 1-8 \AA\ SXR flux. This feature is not clear in the WL curves obtained with the traditional method.

It is noteworthy that, using HMI observational data, \citet{Mravcova2017} also proposed a method of identifying WLFs based on the continuum light curves. They performed a fifth-degree polynomial fit to the light curve at each location, using the fit result as the background to be subtracted. A threshold, $\sigma_m$, was then set to calculate the number of identified WL kernels. They assumed that the area of WL kernels is very small compared to the field of view. As the threshold increases, the number of identified WL kernels decreases rapidly and eventually stabilizes. The $\sigma_m$ value at this critical transition point was selected as the final threshold. 

Since different WLFs vary in radiative intensity and evolutionary trend, this method requires an individually determined threshold for each event. The identification of the critical point may be influenced by the chosen range of thresholds and the step size used in the calculation, potentially leading to different  $\sigma_m$. Additionally, during background subtraction, the use of a fifth-degree polynomial fit may treat some gradual WL emissions as part of the background. Consequently, such signals could be inadvertently removed, leading to a loss of information.

Furthermore, in their identification process, any WL kernel consisting of fewer than 60 pixels was discarded. This criterion appears questionable. \citet{Jess2008} reported a C2.0 WLF with a kernel diameter of only approximately 300 km, which corresponds to just a single pixel at HMI's resolution. Similarly, \citet{Song2020} reported a compact C2.3 WLF with a kernel spanning only about 10 pixels. This implies that their method would fail to detect these small, weak, and compact WLFs, and in fact their study mainly focused on flares with a GOES class above M5.0.
  
In contrast, our identification method independently authenticates the light curve at each individual location. This method is grounded in a clear physical picture, and is straightforward to implement. It allows for the application of a unified threshold (e.g. $(I_{fm}-\frac{1}{N}\sum_{i=1}^{N} I_{0i})\ge6\sigma$) and is robust against the effects of flare size, the intensity of WL emission, and the area of the WL kernels. As will be demonstrated later, this approach successfully extracts detailed WL evolution information from a X2.2 WLF, while also exhibiting excellent detection capability for very weak C-class WLFs in the active region NOAA 12887.

%------------------------------------------------------------------
\subsection{The usage of HMI continuum data for WLFs} \label{Sec3.6}

HMI continuum observations are highly valuable in the statistical study of WLFs \citep[e.g.][]{Kuhar2016, Mravcova2017, Watanabe2017, Song2018c, Jing2024, Cai2024, Li2024}. However, it is should be noted that this continuum was synthetically constructed from filtergrams taken at six points along the observed spectral line of Fe {\scriptsize I}  6173 \AA, using the following formula: 
\begin{equation}
\emph{$I_{c}=\displaystyle\frac{1}{6}~\displaystyle{\sum_{j=0}^5}~[I_j+I_d~exp(-\frac{(\lambda-\lambda_0)^2}{\sigma^2})]$}\label{equation1}
.\end{equation}
Here $\lambda_0$, $\sigma$, and $I_d$ are the rest wavelength, line width, and line depth, estimated by taking six sampling points ($I_j$) across the Fe {\footnotesize I} line of 6173 \AA~ \citep{Couvidat2012}.

Some studies suggest that during quiet periods, the constructed continuum agrees reasonably well with the true continuum \citep[e.g.][]{Svanda2018}. However, during flares, the spectral line undergoes significant perturbations, leading to a considerably larger deviation between the constructed and the true continuum \citep[e.g.][]{Svanda2018, Granovsky2025}. Furthermore, simulations by other researchers have indicated that radiation enhancement originating purely from the line core can, through the construction process, manifest as an apparent enhancement in the synthesized continuum, even in the absence of a true continuum enhancement \citep[e.g.][]{Hong2018, Mravcova2017}.

\begin{figure}[!ht]
\centerline{\includegraphics[trim=0.0cm 0.8cm 0.0cm 0.0cm, width=0.52\textwidth]{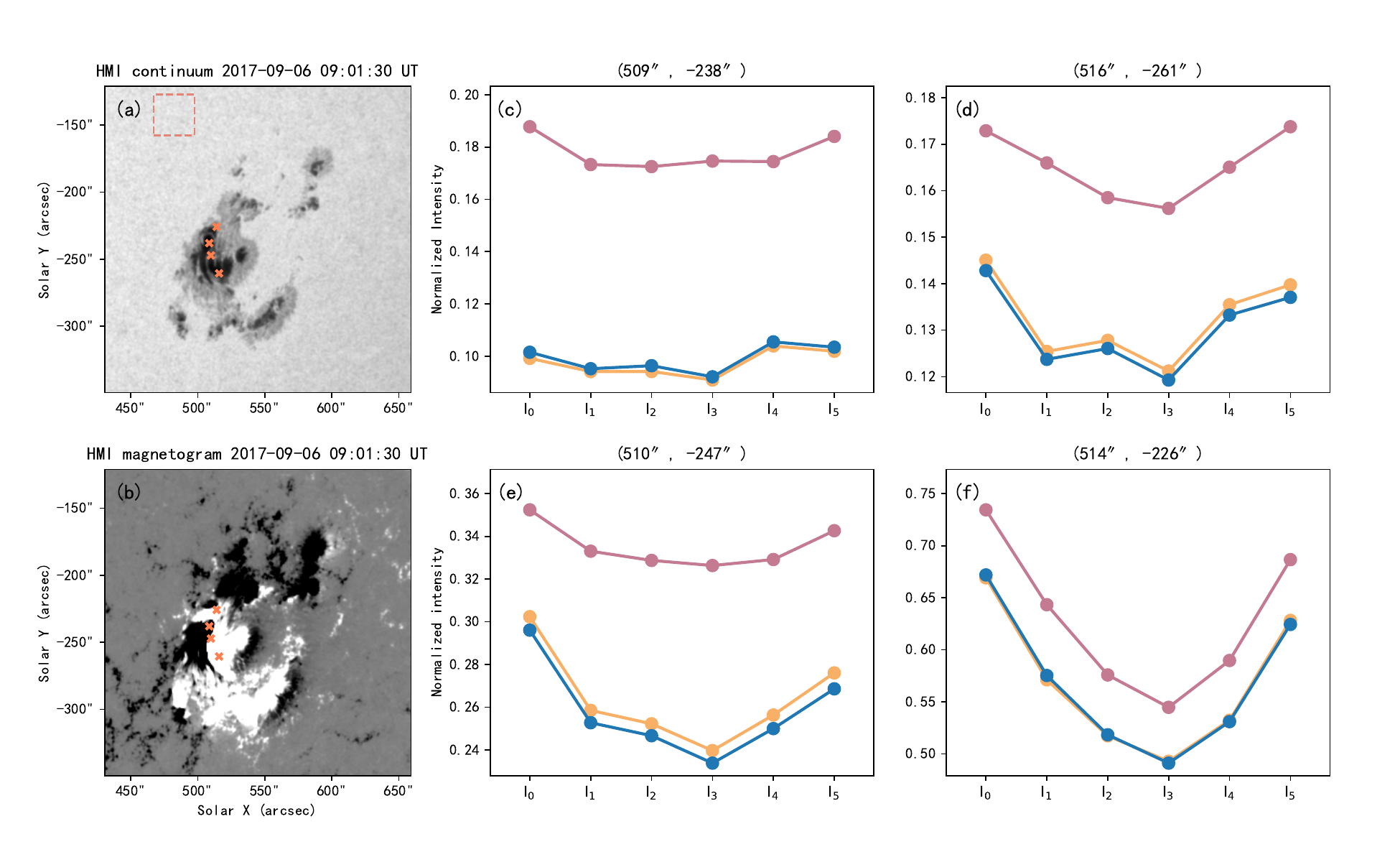}}
\caption{Spectral profiles of HMI Fe {\scriptsize I}  6173 \AA\ at four different positions.  (a) and (b): HMI continuum intensity map and LOS magnetogram, respectively. The orange `$\times$' marks the four positions. (e) and (f): HMI Fe {\scriptsize I}  6173 \AA\ spectral profiles for the four positions at three different times, i.e. 08:48 UT (yellow), 09:00 UT (blue), and 09:12 UT (purple). 08:48 UT and 09:00 UT are before the flare, while 09:12 UT is around the peak of flare. All the spectral profiles are normalized ($I_i/I_q$, $i=0, 1, 2, 3, 4, 5$) by the average continuum intensity ($I_q$) in the quiet region marked by the dashed brown box in panel (a).}
\label{fig8}
\end{figure}

Accordingly, we compared the spectral profiles at four different locations during the flare with those before the flare (Fig. \ref{fig8}). It can be observed that at the two pre-flare times, the radiation intensity at all wavelength positions of the profile was similar or exhibited only minor variations. In contrast, during the flare, the radiation at all wavelength positions increased significantly. This allows us to rule out the possibility that the enhancement originates solely from the line core or a few nearby points. 

This strong enhancement across the entire spectral profile should be closely related to the WLF \citep{Granovsky2025}. However, this does not necessarily mean that the constructed continuum during the flare represents the true continuum. As long as we can confirm a great change in the spectral radiation during the flare compared to the pre- and post-flare periods, it can, to some extent, indicate the occurrence of a WLF. Therefore, for the identification of WLFs, the relative change in radiation intensity is of greater importance. This can be implemented by adopting a sufficiently high threshold, such as the $(I_{fm}-\frac{1}{N}\sum_{i=1}^{N} I_{0i})\ge6\sigma$ criterion used in this study (Fig. \ref{fig3}).

It is important to note that the characteristics of HMI spectral line perturbations may differ significantly across flares of varying classes and types. Simple theoretical modelling may be insufficient to provide a detailed description of these differences. A comprehensive understanding of all the details will require future comparisons with more authentic and reliable observations of the true continuum. This endeavour, however, falls beyond the scope of the present study, and we plan to address it specifically in a future work.

%===============================================================
\section{WL emissions in the X2.2 flare on September 6, 2017} \label{Sec4} 

\subsection{WL emissions associated with the flare evolution} \label{Sec4.1}

\begin{figure}[!ht]
\centerline{\includegraphics[trim=0.0cm 1.1cm 0.0cm 0.0cm, width=0.5\textwidth]{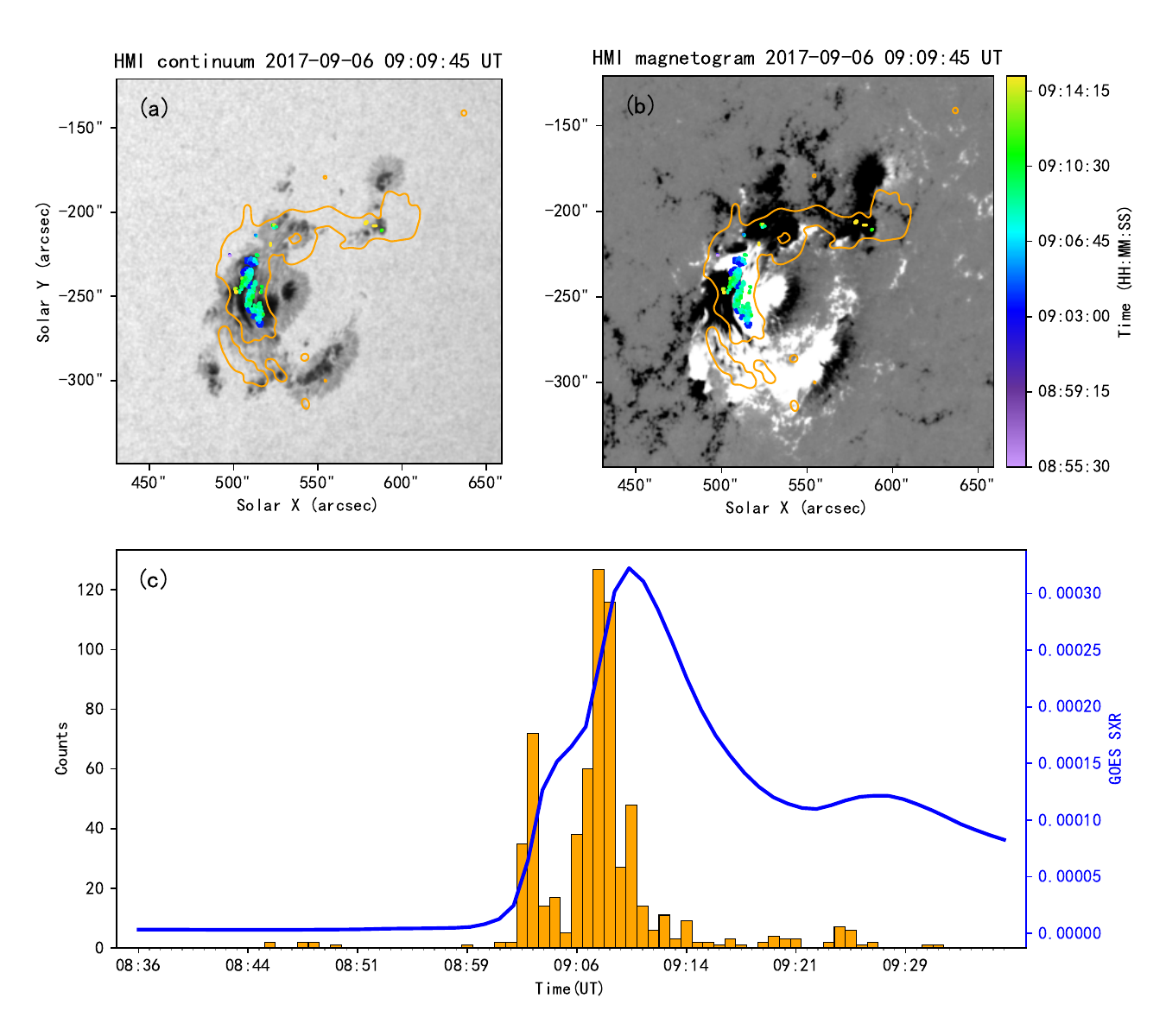}}
\caption{Temporal and spatial evolution of the WL emissions in the X2.2 flare on September 6, 2017. (a) and (b): HMI continuum intensity map and LOS magnetogram, respectively. The orange contours mark the flare region. The marked points represent the qualified WL pixels identified by our method, and the colour of the points indicates the different peak times of the WL flux at each point. (c): Histogram of the WL pixels with different peak times. The blue light curve is GOES SXR 1-8 \AA\ flux.}
\label{fig9}
\end{figure}

Figure \ref{fig9} shows the relationship between the WL emissions and the evolution of this X2.2 flare. Panels (a) and (b) present the spatial distributions of the identified WL pixels with different peak times. We see that there are mainly two peak times for these WL pixels: one is around 09:03 UT and the other is around 09:10 UT. It is also clearly seen that the number of WL pixels with the latter peak time is much larger than those with the peak time around 09:03 UT. These WL pixels are all distinctly distributed within the flare ribbon on both sides of the PIL. We find that these WL pixels with the peak times around 09:03 UT are located at the edge of the entire WL emission region, far from the PIL. In contrast, those WL pixels with the peak times around 09:10 UT are concentrated near the PIL.

Figure \ref{fig9}c shows the statistical results of WL pixels with different peak times and their relationship with the GOES SXR 1-8 \AA\ flux. As is seen from the GOES SXR flux, the rising phase of the flare consists of two stages, corresponding to the periods around 09:03 UT and 09:10 UT, respectively. Moreover, a late phase appears at 09:27 UT after the flare's main decay phase. We see a large number of WL pixels appear in the first stage of the rising phase (around 09:03 UT). However, the second stage of the rising phase (around 09:10 UT) contains the largest number of WL pixels. The GOES SXR flux increases rapidly during the two stages, but slows down in the brief interval between the two stages. After the flare peak, i.e. the decay phase, we see that the number of WL pixels decreases sharply. Notably, the rising stage of the late phase (around 09:24 UT) corresponds to an obvious increase in the number of WL pixels, but the value is much smaller than that in the two stages of the flare's rising phase.

This corresponding relationship indicates that WL pixels mainly appear during the flare's heating phase rather than the cooling phase. Additionally, we find the WL pixels appear throughout nearly the entire flare process, suggesting continuous heating in the lower atmosphere throughout the flare, albeit with varying amplitudes.

%------------------------------------------------------------------
\subsection{Types of WL curves and their characteristics} \label{Sec4.2}

We find that the WL curves identified exhibit diverse characteristics, indicating that the evolution of WL emission varies within different regions. Such differences likely result from the coupling between the flare heating process and the local gradual evolution. Based on the intensity, duration of WL emission pulses during the flare, and the characteristics of background variations, these WL curves can generally be classified into eight categories.

\begin{figure*}
\centerline{\includegraphics[trim=0.0cm 1.6cm 0.0cm 0.0cm, width=0.98\textwidth]{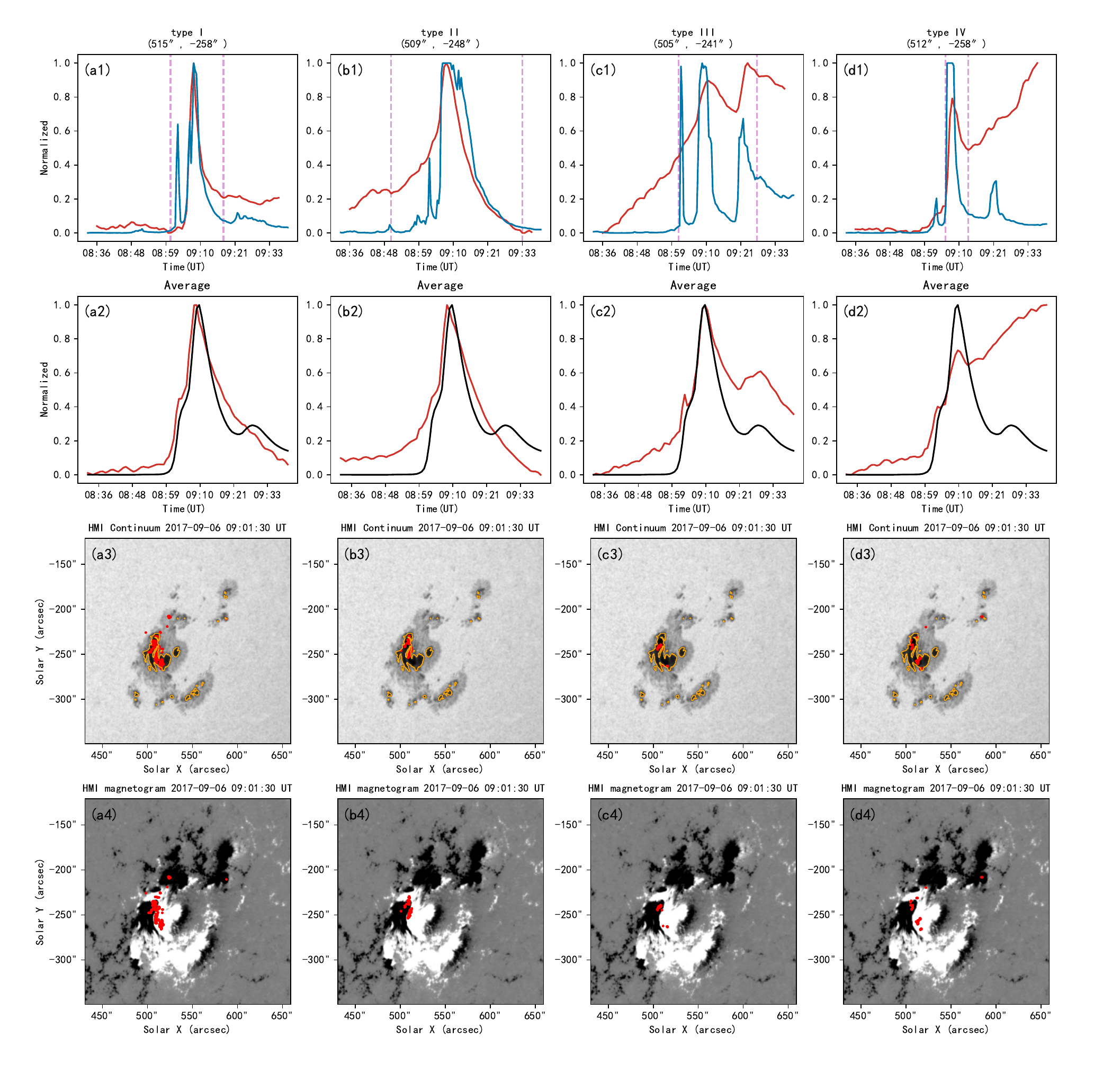}}
\caption{Type-I to type-IV WL emission light curves and their spatial distributions. (a1)-(a4): Light curves and spatial distribution for type-I light curves, which show a rapid enhancement with a short duration. (a1): WL emission light curve (red) at the position of ($555^{\prime\prime}$, $-258^{\prime\prime}$). The blue is the AIA 1600 \AA\ flux at the same position. The vertical purple lines mark the start and end of the WL enhancement during the flare at the position. (a2): Average WL emission light curve (red) for this type of WL pixels. The black represents the GOES SXR 1-8 \AA\ flux. (a3) and (a4): HMI continuum image and LOS magnetogram before the flare, respectively. The red marks indicate the position of the type-I WL pixels. The orange contours in panel (a3) denote the boundary between the sunspot's umbra and penumbra. (b1)-(b4): Type-II light curves and their spatial distribution, which exhibit a slow growth phase followed by rapid growth with a long duration. (c1)-(c4): Type-III light curves and their spatial distribution, which display two consecutive peaks for the WL emission flux.  (d1)-(d4): Type-IV light curves, which are associated with a continuous upward trend in the background.}
\label{fig10}
\end{figure*}

\begin{figure*}
\centerline{\includegraphics[trim=0.0cm 1.6cm 0.0cm 0.0cm, width=0.98\textwidth]{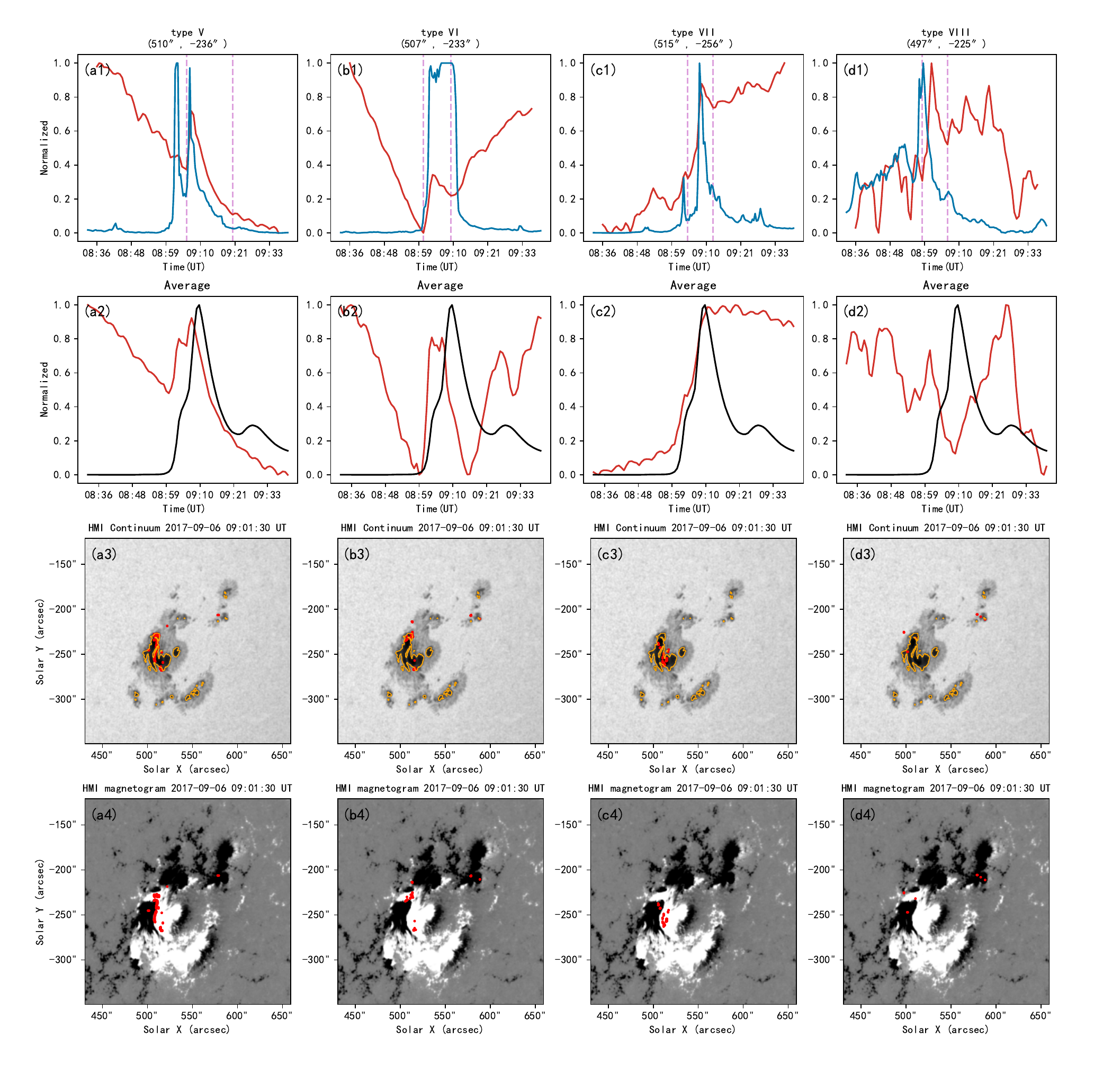}}
\caption{Similar to Fig. \ref{fig10} but for the type-V to type-VIII WL emission light curves. Type-V light curves show a continuous downward trend in the background (a1-a4). Type-VI light curves show a reversal of the background trend before and after the WL emission (b1-b4). Type-VII light curves are associated with a step-like change in the background trend  (c1-c4), while the type-VIII WL emission light curves show a cluttered background (d1-d4).}
\label{fig11}
\end{figure*}

Figures \ref{fig10} and \ref{fig11} show the WL curves and their corresponding spatial distribution for type I-IV and type V-VIII, respectively. Type-I WL curves show a very rapid enhancement with a short duration. The WL pixels exhibiting this type of light curve are widely distributed across the flare ribbon. Type-II WL curves exhibit a slow growth phase followed by rapid growth phase. The duration for WL emission is also much longer than for type I, which even exceeds 30 minutes. In addition, the type-II WL pixels appear to be mainly distributed within the flare's core region.

\begin{table*}[htbp]
\centering
\caption{Statistics of the eight types of WL emission light curves.}
\label{tab1}
{\fontsize{8.0}{8.8}\selectfont
\begin{tabular}{ccccccccccc}
\hline
\hline
Type &Feature description&Number of pixels&Percentage\\
\hline
        I & rapid enhancement with a short duration & 203 & 35.37\%  \\
        II & a slow growth phase followed by rapid growth with a long duration & 79 & 13.76\%  \\
        III & two consecutive peaks for the WL emission flux & 27 & 4.70\% \\
        IV &  associate with a continuous upward trend in the background & 56 & 9.76\%  \\
        V & show a continuous downward trend in the background & 123 & 21.43\% \\
        VI & show a reversal of the background trend before and after the WL emission & 34 & 5.92\%  \\
        VII & associate with a step-like change of the background trend & 43 & 7.49\%  \\
        VIII & show a cluttered background & 9 & 1.57\%  \\
        \hline
        Total &  & 574 & 100\% \\                                                                                                       
\hline
\end{tabular}
}
\end{table*}

Type-III WL curves display two consecutive peaks, indicating WL radiation events occurring in two consecutive time periods. It is interesting to see that these WL curves are highly coincident with the variations in the GOES SXR flux (Fig. \ref{fig10}c2). The second peak also precisely corresponds to the late phase of the flare. For this type of WL curves, some pixels are distributed in the umbra regions and others are located around the PIL.

For the type-IV and type-V WL curves, the main difference is their background trends. The background for the type-IV WL curves exhibits a continuously rising trend, while the background for the type-V WL curves displays a systematically declining trend. The type-IV WL pixels are mainly located in the umbra regions, while the type-V WL pixels are predominantly concentrated along the PIL.

The type-VI WL curves are characterized by a distinct reversal in background trends before and after the WL emission. We see a continuous decrease background trend before WL radiation, which turns into a continuous increase after the WL radiation. Notably, the peaks of this type WL curves are mostly concentrated around 9:03 UT, indicating that this type of WL emission occurs mainly during the first stage of the flare. The WL pixels of this type are primarily distributed at the boundaries of sunspots and network field regions. 

Type-VII WL curves present a background that rises continuously before the peak. After the WL radiation ends, the background decrease is not significant; instead, it remains oscillatory near the peak. This type of WL pixel is widely distributed in the umbral and penumbral regions of sunspots. Type-VIII WL curves show a cluttered background. There are far fewer pixels for this type of WL curves and they are sporadically distributed in the network fields and quiet regions.

Table 1 gives a simple statistic of the eight types of WL curves. We identify 574 WL pixels in total using our method. The type-I WL pixels are the most numerous and the type-VIII WL pixels are the least. It is noteworthy that regardless of how complex the WL curve variations are, our method remains effective in identifying them.
%------------------------------------------------------------------
\subsection{Statistics of WL enhancements and durations within different regions} \label{Sec4.3}

\begin{figure}[!ht]
\centerline{\includegraphics[width=0.55\textwidth]{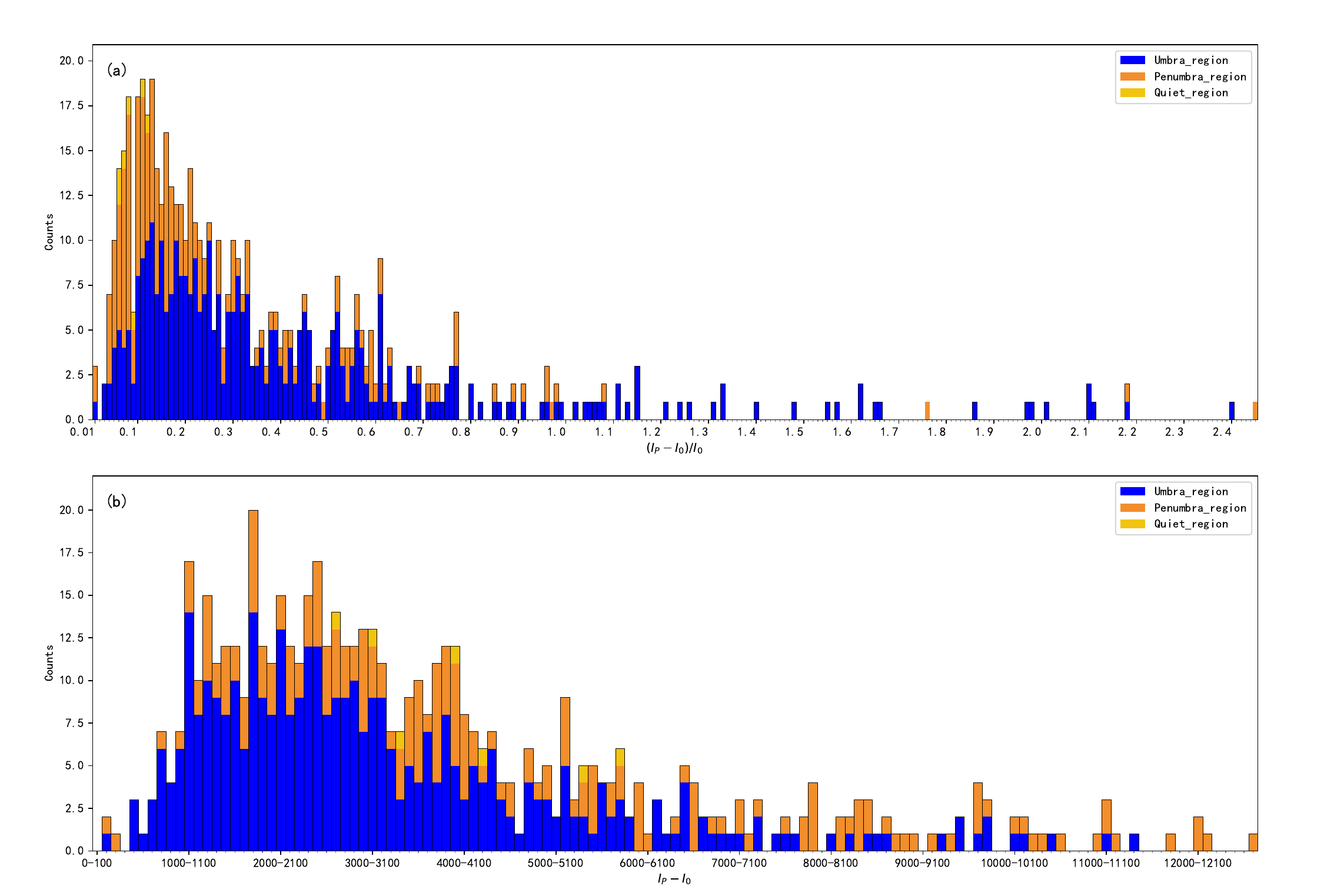}}
\caption{Statistical distribution of relative WL enhancement (a) and absolute WL enhancement (b) in different regions of the Sun. Blue, orange, and yellow colours correspond to the umbra, penumbra, and quiet regions, respectively.}
\label{fig12}
\end{figure}

Figure \ref{fig12} presents the statistical distribution of relative and absolute WL enhancements for these WL pixels in three regions, i.e. umbra, penumbra, and quiet regions. The relative WL enhancement and the absolute WL enhancement are defined as $(I_p-I_0)/I_0$ and $I_p-I_0$, respectively. The weakest WL enhancement we detected is 1\%, which is currently the lowest known WL enhancement (also see Fig. \ref{fig6}). Notably, we see that there are still a considerable number of WL pixels with a WL enhancement  below 8\%, indicating that the threshold of 8\% set by previous studies \citep[e.g.][]{Song2018b, Song2018c, Jing2024, Li2024} is relatively high.

The ranges of relative WL enhancements for WL pixels in different regions are different. Since different regions have inherently different levels of brightness, the calculated relative intensity enhancements will differ significantly. The relative WL enhancements in quiet regions are relatively small, ranging from $\sim6\%$ to $\sim12\%$. The relative WL enhancements in penumbral regions range from $\sim4\%$ to $\sim30\%$, while the umbral regions exhibit the largest range of relative WL enhancements, from $\sim3\%$ to $\sim80\%$. The relative WL enhancements for most WL pixels are within 100\%, with only a few exceeding this value. However, for the  absolute WL enhancement, there is no significant difference among the WL pixels in these regions, with them ranging from $\sim500$ DN to $\sim12000$ DN. It is also seen that a large number of WL pixels are concentrated in the umbral regions, with only a small number in penumbral and quiet regions. 

\begin{figure}[!ht]
\centerline{\includegraphics[width=0.5\textwidth]{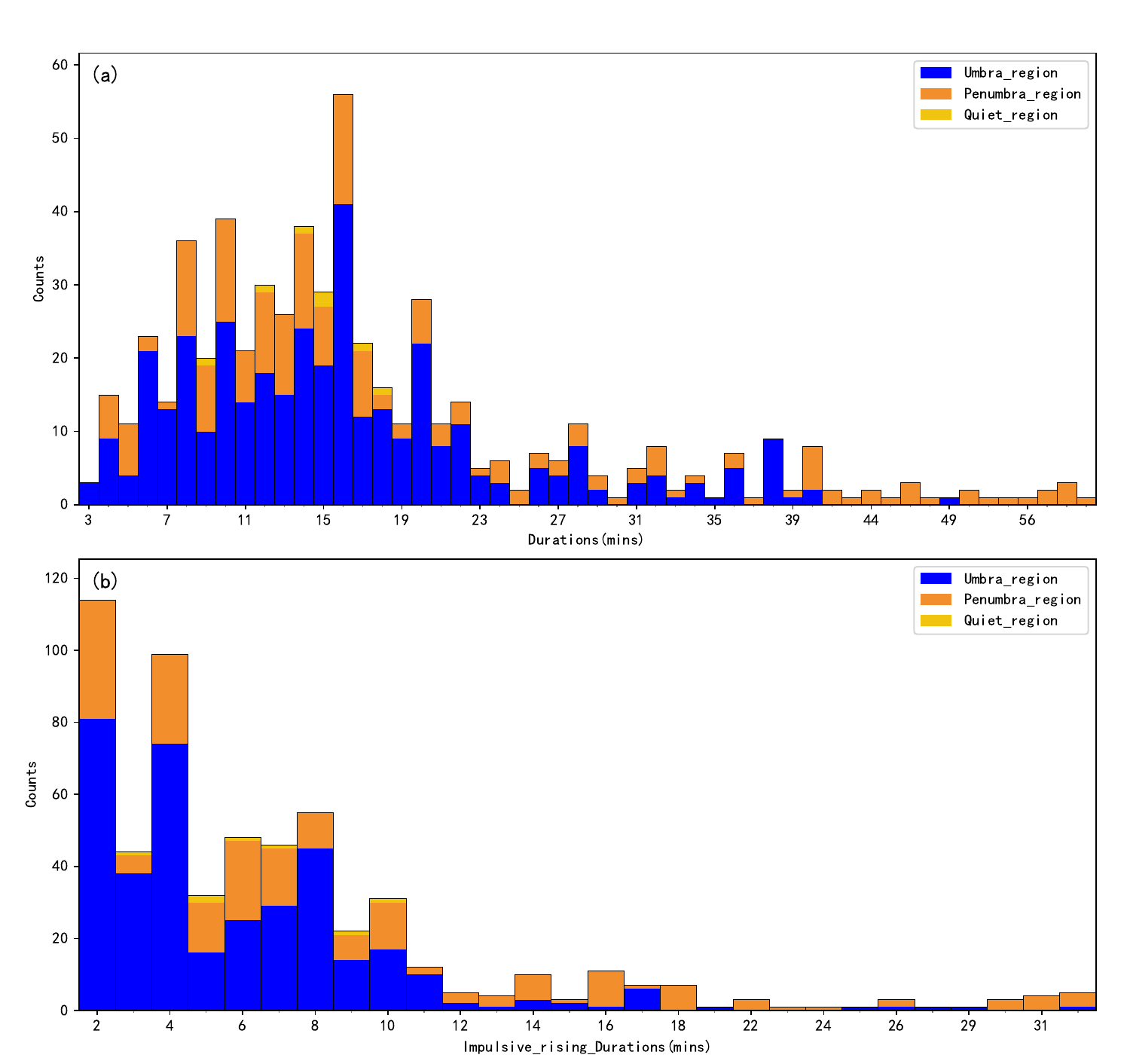}}
\caption{Similar to Fig. \ref{fig12}; the statistical distributions of the WL durations (a) and WL impulsive rising phase durations (b) for different regions.}
\label{fig13}
\end{figure}

Figure \ref{fig13} shows the statistical distributions of the durations of WL emissions and the durations of the rising phase of WL emissions in these three regions. There is no significant difference among the umbra, penumbra, and quiet regions for these two kinds of durations. The durations of WL emissions are mainly concentrated between $\sim4$ and $\sim30$ minutes, with a few WL pixels reaching $\sim40$ minutes. The duration for the rising phase is mainly concentrated between $\sim2$ and $\sim11$ minutes, accounting for approximately half of the WL emission durations, and a small number of WL pixels have a rising phase duration of around 20 minutes.

%------------------------------------------------------------------
\subsection{Relationship between WL enhancement, magnetic field strength, and duration} \label{Sec4.4}

\begin{figure}[!ht]
\centerline{\includegraphics[trim=0.0cm 1.5cm 0.0cm 1.0cm, width=0.52\textwidth]{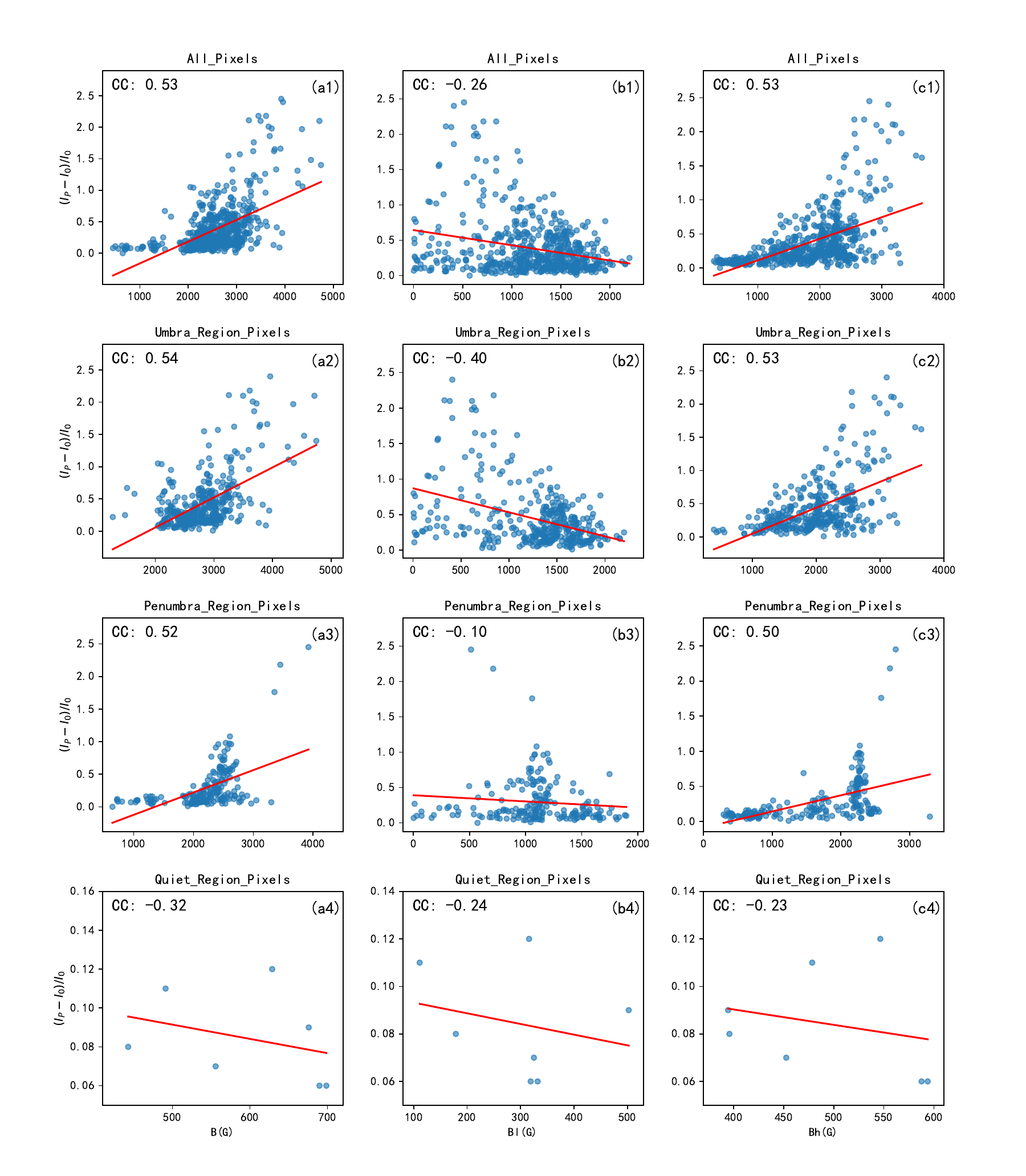}}
\caption{Scatter plots of WL enhancements ($(I_p-I_0)/I_p$) and magnetic field strengths in different regions. (a1) to (a4): Vector magnetic field (B). (b1) to (b4): LOS magnetic field (Bl). (c1) to (c4): Horizontal magnetic field (Bh).  }
\label{fig14}
\end{figure}

Figure \ref{fig14} shows the relationship between the relative WL enhancements and magnetic field strengths. Panels (a1), (b1), and (c1) present the relationships between the relative WL enhancements of all WL pixels and the total magnetic field strength (B), LOS field strength (Bl), and horizontal magnetic field strength (Bh), respectively. It seems that the relative WL enhancements correlate with both the total magnetic field strength and horizontal field strength, though the corresponding correlation coefficients are both only 0.53. 

\begin{table*}[!ht]
\centering
\caption{All the flares in active region NOAA 12887.}
\label{tab2}
{\fontsize{8.5}{9.8}\selectfont
\begin{tabular}{ccccccccccc}
\hline
\hline
Number & Date & Goes peak & Goes class & Location & WLF & WL pixels & WL peak & $\Delta I_{wl}$ (mean) & $\Delta I_{wl}$(max) \\ 
\hline
        1 & 2021-10-26 & 00:24 & C1.9  & S28E30 & Yes & 7 & 00:22 & 5\% & 8\% \\
        2 & 2021-10-27 & 03:56 & C1.0   & S25E16 & Yes & 7 & 03:54 & 9\% & 18\% \\
        3 & 2021-10-27 & 21:59 & C1.1  & S26E05 & Yes & 13 & 21:59 & 7\% & 17\% \\
        4 & 2021-10-27 & 22:44 & C1.1  & S28E07 & No & --- & --- & --- & --- \\
        5 & 2021-10-27 & 07:10 & C1.5 & S27E14  & No & --- & --- & --- & --- \\
        6 & 2021-10-27 & 06:02 & C8.5  & S26E05 & Yes & 6 & 05:59 & 8\% & 13\% \\
        7 & 2021-10-27 & 22:22 & C1.6  & S30E11 & Yes & 10 & 22:20 & 7\% & 9\% \\
        8 & 2021-10-27 & 08:49 & C1.7  & S30E11 & Yes & 20 & 08:48 & 7\% & 12\% \\
        9 & 2021-10-27 & 12:19 & C2.3  & S26E14  & Yes & 13 & 12:14 & 5\% & 9\% \\
    10 & 2021-10-27 & 09:54 & C2.8  & S27E12 & Yes & 17 & 09:57 & 5\% & 10\% \\
    11 & 2021-10-28 & 10:00 & C1.1 & S31E00  & No & --- & --- & --- & --- \\
    12 & 2021-10-28 & 01:38 & C1.2 & S26E03  & Yes & 4 & 01:35 & 7\% & 9\% \\
    13 & 2021-10-28 & 21:10 & C1.5 & S31W05 & Yes & 4 & 21:09 & 4\% & 6\% \\
    14 & 2021-10-28 & 13:21 & C3.3 & S26W03 & Yes & 20 & 13:21 & 7\% & 13\% \\
    15 & 2021-10-28 & 13:59 & C3.8  & S31E01 & No & --- & --- & --- & --- \\
    16 & 2021-10-28 & 07:39 & M1.4 & S31E02  & Yes & 76 & 07:40 & 8\% & 15\% \\
    17 & 2021-10-28 & 10:28 & M2.2 & S30E00  & Yes & 94  & 10:26 & 8\% & 37\% \\
    18 & 2021-10-28 & 15:35 & X1.0  & S28W01 & Yes & 37 & 15:30 & 8\% & 18\% \\
    19 & 2021-10-29 & 07:50 & C1.1 & S28W18  & No & --- & --- & --- & --- \\
    20 & 2021-10-29 & 13:30 & C2.5 & S30W16  & Yes & 9 & 13:30 & 6\% & 8\% \\
    21 & 2021-10-30 & 09:22 & C1.0 & S29W28  & No & ---  & --- & --- & --- \\
    22 & 2021-10-30 & 21:29 & C1.0 & S30W24  & No & --- & --- & --- & --- \\
    23 & 2021-10-30 & 09:36 & C1.1 & S29W28  & No & --- & --- & --- & --- \\                                                                                              
\hline
\end{tabular}
}
\begin{tablenotes}
        \tiny
        \item[1] $\Delta I_{wl}$ refers to the WL enhancement calculated by $(I_p-I_0)/I_0$. Here $I_p$ and $I_0$ refer to the HMI continuum intensities at the peak of WL flux and before its appearance, respectively.
      \end{tablenotes}
\end{table*}

We further examined the relationships between the relative WL enhancements and magnetic field strength within different regions, i.e. the umbral, penumbral, and quiet regions.  The correlations for the umbral region are similar to those for all the WL pixels (see panels a2, b2, c2, a1, b1, and c1). The relative WL enhancements in the penumbral region also appear to correlate better with the total magnetic field strength and horizontal field strength, with the correlation coefficients of 0.52 and 0.50, respectively. The quiet region contains very few WL pixels with relatively weak WL enhancements. Their WL enhancements show no obvious correlation with the strengths of the total magnetic field, LOS field, or horizontal field.

\begin{figure}[!ht]
\centerline{\includegraphics[trim=0.0cm 2.0cm 0.0cm 1.0cm, width=0.52\textwidth]{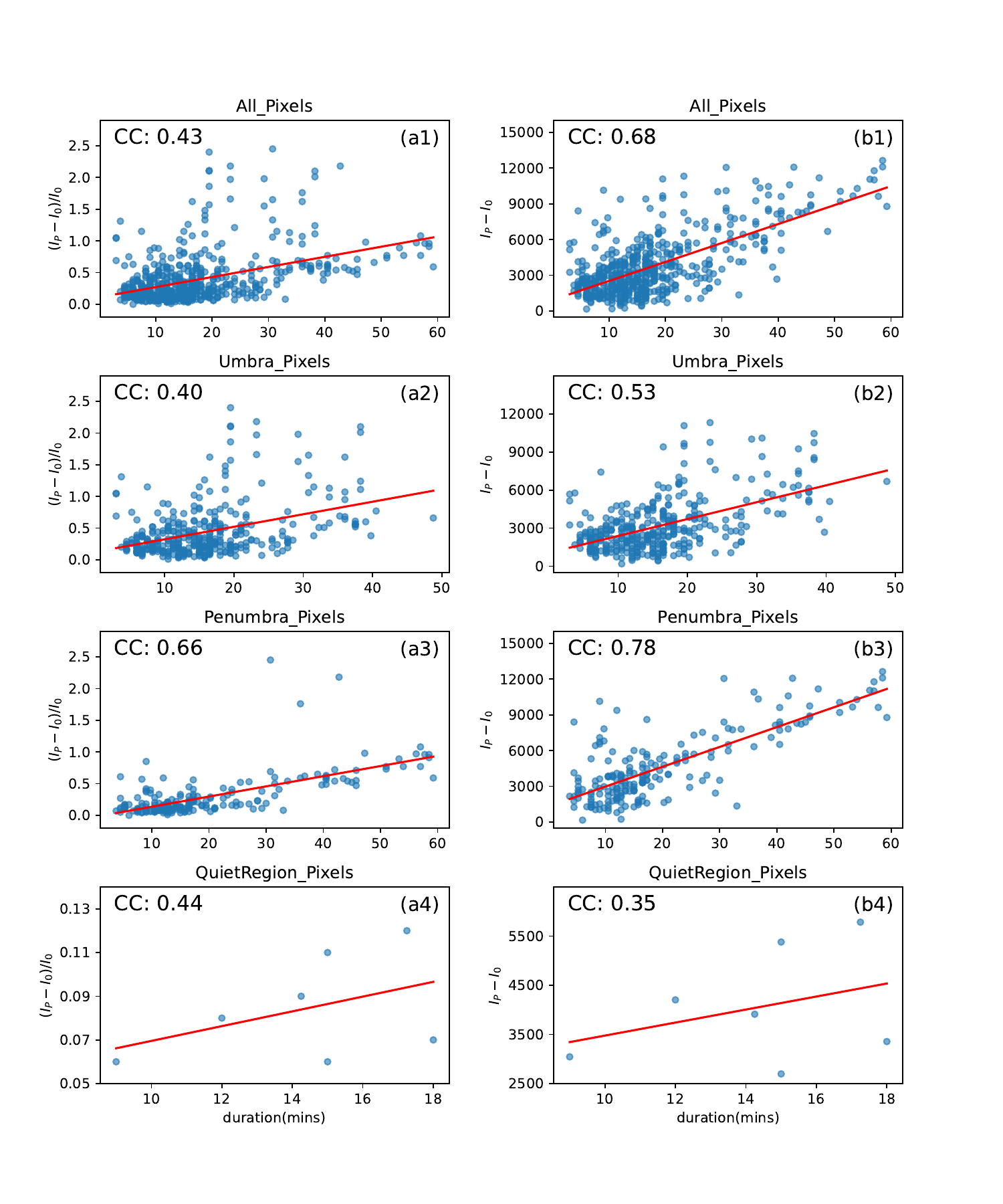}}
\caption{Scatter plots of WL enhancement and WL duration for different regions. (a1) to (a4): Relative WL enhancement ($(I_p-I_0)/I_p$).  (b1) to (b4): Absolute WL enhancement ($I_p-I_0$).}
\label{fig15}
\end{figure}

Interestingly, we note that in the umbral WL regions, the horizontal component of the magnetic field is significantly stronger than the LOS component (see panels b2 and c2), indicating rotating or shearing motions in the sunspot region. Meanwhile, for the penumbral WL regions we see a large part of pixels with the relative WL enhancement greater than $\sim$20\% exhibit a good correlation with the magnetic field strength (panel a3). The corresponding horizontal fields for these pixels are also the strongest in the penumbra (panel c3). 

In fact, this X2.2 flare occurred in an extremely complex active region, NOAA 12673, which is characterized by the emergence of multiple pairs of sunspots, with opposite polarities approaching, rotating, and shearing against each other, forming a highly elongated PIL. These WL pixels are predominantly distributed on both sides of the PIL, suggesting that the occurrence of this WLF is closely related to such strong rotating and shearing motions. It is worth noting that approximately three hours after this event, the largest flare (X9.3) of Solar Cycle 24 occurred in the same active region. Some studies have indicated that prior to this major eruption, multiple magnetic flux ropes had formed along the PIL as a result of the strong rotating and shearing motions of the magnetic field \citep[e.g.][]{Yang2017, Hou2018}.

\begin{figure*}
\centering
\includegraphics[trim=0.0cm 0.6cm 0.0cm 0.0cm,width=0.9\textwidth]{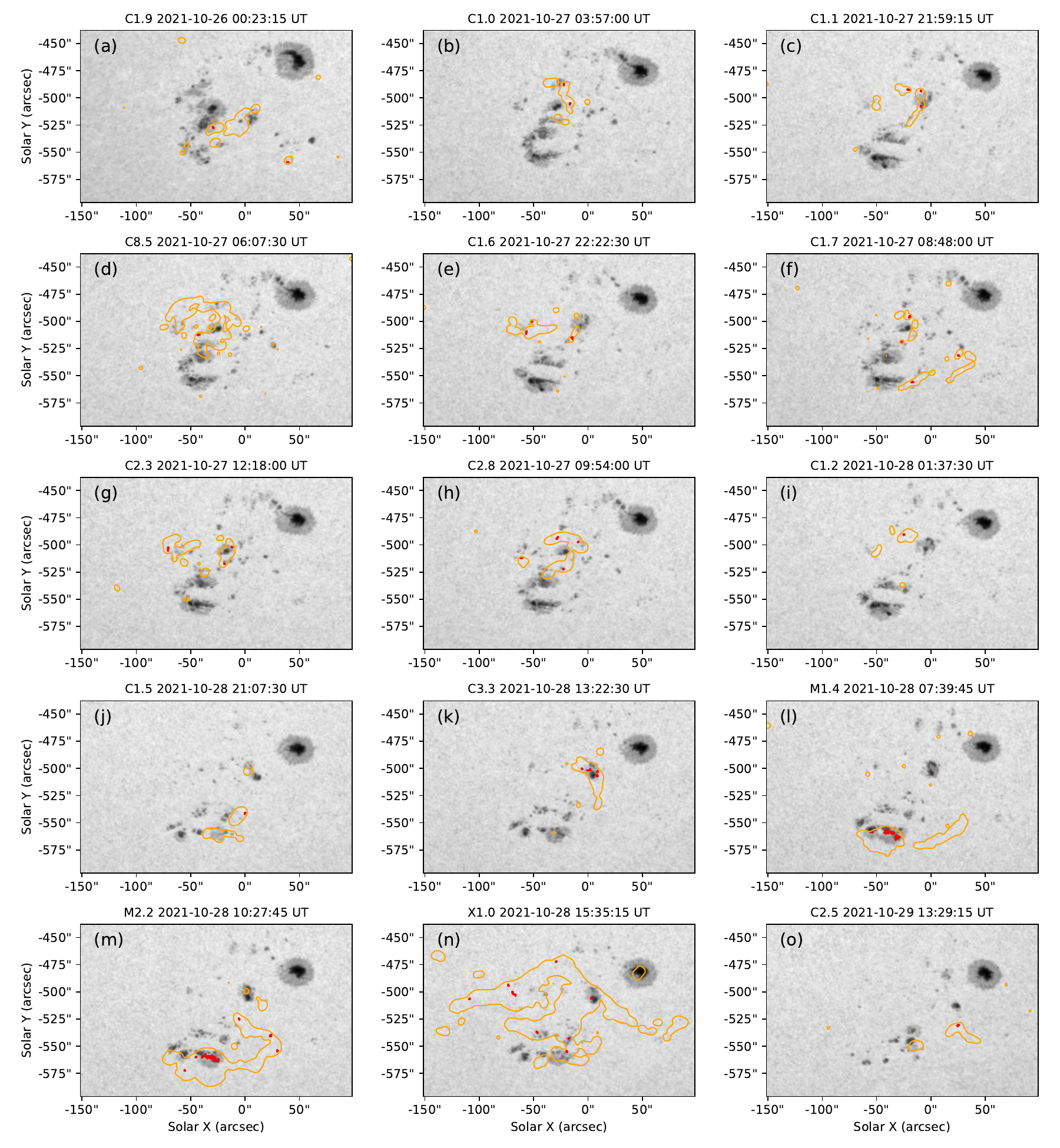}
\caption{Fifteen WLFs identified in the active region of NOAA 12887. The orange contours outline the flare region from AIA 1600 \AA\ observations. The red areas indicate the corresponding WL emission regions.}
\label{fig16}
\end{figure*}

\begin{figure*}
\centering
\includegraphics[trim=0.0cm 2.5cm 0.0cm 0.0cm,width=0.9\textwidth]{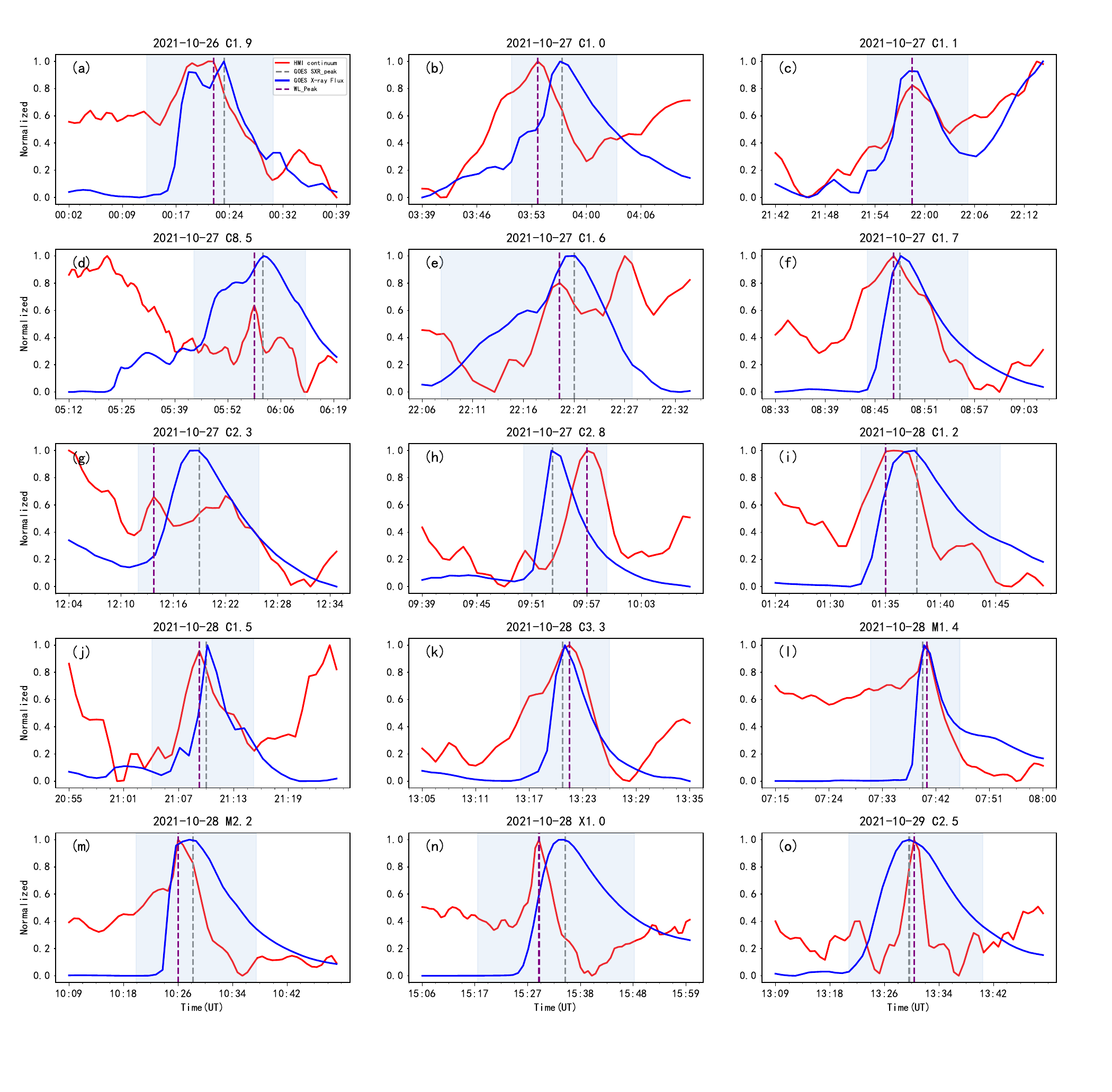}
\caption{Light curves for 15 WLFs observed in the active region of NOAA 12887. The red curve in each panel is the average of WL fluxes for all the identified WL pixels in a flare. The blue curve is the GOES SXR 1-8 \AA\ flux. The vertical purple and grey lines indicate the peak of the WL emission flux and the peak of GOES SXR 1-8 \AA\ flux, respectively. All light curves are normalized. The light blue shaded areas indicate the flaring periods.}
\label{fig17}
\end{figure*}

Figure \ref{fig15} presents the relationships between the relative WL enhancement, absolute WL enhancements, and WL emission durations for WL pixels within different regions. We see that both relative and absolute WL enhancements are highly correlated with WL emission durations for all the WL pixels (panel a1 and b1). The correlation coefficient for absolute WL enhancements reaches 0.68. The relative and absolute WL enhancements in the umbral, penumbral, and quiet regions all show a positive correlation with WL emission durations. Moreover, the correlation of absolute WL enhancements with WL emission durations is generally higher than that for relative WL enhancements. Notably, the penumbral region has the highest correlations, with correlation coefficients of 0.66 for relative WL enhancements and 0.78 for absolute WL enhancements. These strong correlations between them are reasonable. The longer the WL emission durations, the longer the heating process lasts, the more energy is transported to the deeper atmospheric layers, and correspondingly the higher the relative and absolute WL enhancements appear.

%========================================================================================
\section{WLFs in the active region of NOAA 12887} \label{Sec5} 

In order to verify the effectiveness of the proposed WLF identification method, we applied it to detect the WLFs in the active region of NOAA 12887. This active region, with a middle level of activity, produced 23 flares during its passage across the solar disc, including 1 X-class, 2 M-class, and 20 C-class flares. Among the C-class flares, most are very small flares. Only one exceeded C5.0, while the rest were below C5.0. Table 2 lists all of the flares that occurred in the active region of NOAA 12887.

Using our method, we identified 15 WLFs, consisting of 1 X-class, 2 M-class, and 12 C-class. The total occurrence rate of WLFs in this active region is approximately 65\% (15/23). The occurrence rate of WLFs among C-class flares reaches 60\% (12/20). This is the highest occurrence rate of WLFs among C-class flares in all existing statistics. Additionally, we detect the C1.0 WLF, which is the lowest GOES class WLF reported to date. All of this indicates that WL emission is common even in C-class flares, though the WL emission is relatively weak with few pixels. C-class WLFs reported in previous studies are mainly above C5.0 \citep[e.g.][]{ Matthews2003,Cai2024, Xuzhe2025}. Only a few have been reported below C5.0 \citep{Hudson2006, Jess2008, Song2020, Liqiao2024}. Furthermore, the lowest number of WL pixels in a WLF detected in our study is only 4.

The average WL enhancements ($\Delta I_{wl}$ (mean)) of these WLFs are all within 10\%, while the maximum WL enhancement ($\Delta I_{wl}$ (max)) can reach 37\% in the M2.2 flare. The maximum WL enhancements in the X1.0 and the M1.4 flares are 18\% and 15\%, respectively. It is noteworthy that the lowest WLF we detected, i.e. the C1.0 flare, also exhibits a maximum WL enhancement of 18\%.

Figure \ref{fig16} illustrates the flare ribbons and the WL enhancement regions for these 15 WLFs. Except for the two M-class WLFs, which contained a substantial number of WL pixels concentrated in penumbral regions, other WLFs had only a small number of WL pixels scattered across the flare ribbon, and distributed in penumbral, network, or quiet regions. Additionally, the WL emission regions for several C-class flares consist of a single kernel. The three WLFs with the most WL pixels are the M1.4, M2.2, and X1.0, containing 76, 94, and 31 pixels, respectively. The number of WL pixels in the remaining flares was generally lower than 20.

Figure \ref{fig17} shows a comparison between the average WL curves and the GOES soft X-ray 1-8 \AA\ fluxes for these 15 WLFs. It can be observed that the peaks of some WL emission light curves precede those of soft X-rays, some are nearly synchronous, and others lag behind the peaks of soft X-ray flux. This phenomenon indicates that the roles and contributions of thermal and non-thermal mechanisms vary among different WLFs.
%========================================================================================
\section{Summary and discussion} \label{Sec6} 

We propose a new method of identifying WLFs. The basic idea is to decompose each light curve in the flaring region into two parts. One is the slow or even periodic background variation, which is generally caused by the p-mode waves, granulation motion or magnetic field evolution. The other is the rapid pulse-like variation, which is usually caused by the instantaneous heating of the lower atmosphere by energy released from flares, jets, or other magnetic reconnection processes. Based on this, a series of WL radiation pulses are obtained by subtracting the background. Finally, the WL curves are determined by comparing the radiation pulses during the flare with those before and after the flare, and the corresponding pixels are designated as WL pixels. 

This method has the following advantages:
\begin{itemize}
\item[1.] It does not need to set a lower limit or threshold for WL radiation enhancement, but only considers the radiation characteristics of the continuum light curve at each pixel. Therefore, we can detect extremely weak WL emission enhancements (see Fig. \ref{fig6}).

\item[2.]  It fully examines the continuum light curve at each position and will not miss some WL pixels due to the different WL emission peak times. We can also obtain the details of WL radiation evolution in a flare (see Fig. \ref{fig9}).
\end{itemize}

We first applied this method to the X2.2 flare of September 6, 2017 and obtained the following results:
\begin{itemize}
\item[1.] It was found that there were two stages in the rising phase of the flare, and that the WL pixels mainly correspond to two peak times. One is around 09:03 UT and the other is around 09:10 UT. In addition, during the rising stage of the late phase of the flare, there is also a relatively obvious increase in the number of WL pixels. Obviously, the WL enhancements mainly occur during the flare heating phase, corresponding to the rise of the GOES SXR 1-8 \AA\ flux, while WL pixels disappear rapidly during the flare cooling phase, i.e. when the GOES SXR 1-8 \AA\ flux decreases.

\item[2.] The WL pixels exist throughout the entire flare period, but the number of WL pixels and the corresponding intensity of WL enhancements vary in different stages.

\item[3.] The WL curves in this flare can be divided into eight types, which should be caused by the coupling of different heating processes, height of radiation, cooling timescales, and multiple episodes of energy release during the flare with the complex background variations.

\item[4.] The WL pixels are predominantly located in umbral and penumbral regions on both sides of the PIL, where the horizontal magnetic field is significantly stronger than the LOS field, suggesting the presence of sunspot rotation and strong shearing motions around the PIL. Therefore, we infer that the generation of this WLF is closely related to such rotating and shearing motions of the magnetic field within the active region.

\item[5.] The WL enhancement is positively correlated with the durations of WL radiation, indicating that the longer the heating lasts, the greater the intensity of the corresponding WL enhancement.
\end{itemize}
    
To verify the effectiveness of the method, we used it to examine all the flares that occurred in the active region of NOAA 12887, which produced 1 X-, 2 M-, and 20 C-class flares during its passage across the solar disc. We found that:
\begin{itemize}
\item[1.] Among the 23 flares in this active region, there are 15 WLFs, including 1 X-class, 2 M-class, and 12 C-class. The total WLFs occurrence rate is 65\%, and the occurrence rate among C-class flares is 60\%. This is the highest occurrence rate of WLFs in C-class flares reported so far. It indicates that WL radiation may be a common phenomenon, even in C-class flares.

\item[2.] We detected a C1.0 WLF and this is the currently lowest GOES class WLF, demonstrating the extremely high detection capability of our method.

\item[3.] It was found that there are various situations between the peaks of WL radiation and the peaks of GOES SXR 1-8 \AA\ flux: precedence, synchronization, and lag. This indicates that the thermal and non-thermal mechanisms and processes producing WL radiation in different flares are different, and also indicates the complexity of the WLF mechanisms. It should be noted that a WLF may contain multiple WL kernels, each appearing at different phases of the flare and evolving asynchronously, resulting in varying peak intensities. Consequently, the time lag between the peak of the integrated WL light curve and the peak of the GOES SXR flux may vary from event to event. In the future, a detailed analysis of each individual WL kernel will be required to fully uncover the intricate processes of lower atmospheric heating.
\end{itemize}

Through the new WLF identification method and its application, we have found that the evolution of WLF has rich details. The WL radiation is commonly present in general flares, including C-class flares. In particular, we find some very weak C-class flares that below C5.0 also exhibit detectable WL emissions (see Table 2 and Figs. \ref{fig16} and \ref{fig17}). Our findings seem to support the idea that WL radiation may be a universal feature of all solar flares. All these results provide new insights into WLFs. In the future, we shall conduct further in-depth research on WLFs based on this method.

%========================================================================================
\begin{acknowledgements}
This work is supported by National Natural Science Foundation of China (Grant No. 12173049, 12373035, 12473053, 11803002, 11973056), National Key R\&D Program of China (Grant No. 2022YFF0503001, 2021YFA0718601) and Beijing Natural Science Foundation (Grant No. 1222029). The authors thank the SDO and GOES teams for providing the data. SDO is a space mission in the Living With a Star Program of NASA.

\end{acknowledgements}
%========================================================================================

\bibliographystyle{aa}
\bibliography{bibliography}

\end{document}